\documentclass[aps,onecolumn,10pt,superscriptaddress,nofootinbib]{revtex4-2}

\usepackage{amsmath,amssymb}
\usepackage{mathrsfs}
\usepackage{graphicx}
\usepackage{booktabs}
\usepackage{array}
\usepackage{hyperref}

\newtheorem{myproposition}{Proposition}[section]
\newenvironment{proof}{\par\noindent\textit{Proof.}\ }{\hfill$\square$\par}

\hypersetup{
  colorlinks=true,
  linkcolor=blue,
  citecolor=blue,
  urlcolor=blue,
  pdftitle={Curvature-history tensor memory in inflationary cosmology: Component closure, ghost-free dynamics, and CMB BipoSH likelihood},
  pdfauthor={G. M. Rhythm},
  pdfkeywords={Inflationary cosmology, Curvature history, Tensor memory, Statistical anisotropy, Ghost-free nonlocal dynamics, Primordial perturbations, Cosmic microwave background, Bipolar spherical harmonics}
}

\begin{document}

\title{Curvature-history tensor memory in inflationary cosmology: Component closure, ghost-free dynamics, and CMB BipoSH likelihood}
\author{G. M. Rhythm}
\email{rhythm.research.astro@gmail.com}
\thanks{ORCID: 0009-0003-9964-7313}
\affiliation{School of Physics, Damghan University, P.O. Box 3671641167, Damghan, Iran}
\affiliation{Kushtia Zilla School, Kushtia 7000, Khulna, Bangladesh}
\begin{abstract}
We present a generally covariant inflationary model in which curvature history activates a spatial symmetric-tracefree tensor-memory order parameter and produces a scale-dependent quadrupolar departure from primordial statistical isotropy. The curvature source is the four-dimensional Gauss--Bonnet completion of the Kretschmann invariant. The local $(C,\chi)$ system is used only to write the retarded theory in variational form: fixed retarded Cauchy data remove both homogeneous response solutions, so the negative-sign localized combination is not an independent phase-space direction. The projected tensor sector is described by the dressed field $Q_{\mu\nu}=\mathcal E_C\widehat Q_{\mu\nu}$, with $\mathcal E_C=\exp[B(C)\Box_{\widehat Q}/2]$. Its raw-field Hessian is the exact congruence $\mathcal E_C^\dagger(\Box_Q-\mathsf M^2)\mathcal E_C$. Therefore, an invertible dressing preserves both the number of solutions and the sign of the physical symplectic norm even when $C=C(t)$ and $[B(C),\Box_Q]\ne0$.

We derive the full quadratic Hessian for $\zeta$, $\delta C$, $\delta\chi$, the five STF tensor fluctuations, lapse, and shift. Eliminating the ADM constraints and then the retarded response variables gives an exact block Schur complement. Exchange through the tensor-memory sector begins at second order in the weak anisotropic coupling, including the two orientational Goldstone modes. The leading $O(b)$ curvature action is therefore the anisotropic action used for the primordial calculation. For the curvature source, we derive the nonlinear ADM Legendre matrix and the Dirac--Bergmann constraint chain. Along the representative inflationary solution, its determinant stays more than seven orders of magnitude away from the singular threshold, excluding an Ostrogradsky branch. A smooth curvature-dependent phase potential is solved dynamically and gives an approximately constant activated plateau with subdominant memory energy. The finite episode then produces an infrared-regular oscillatory quadrupole and an exact scale-dependent $T/E$ BipoSH covariance. We also give the corresponding cut-sky Gaussian and quadratic maximum-likelihood estimators.
\end{abstract}

\maketitle

\section{Introduction}\label{Xsec1-1}
\label{sec:main_introduction}

Inflation explains the large-scale homogeneity and spatial flatness of the Universe and the nearly scale-invariant spectrum of primordial curvature perturbations. In the minimal picture, a scalar inflaton and the metric fluctuate around an approximately de Sitter background, and the scalar two-point function is statistically isotropic. After the nondynamical gravitational variables are removed, one scalar degree of freedom obeys the Mukhanov--Sasaki equation, and the leading CMB covariance is diagonal in $(\ell,m)$ \cite{bib0011,bib0012,bib0013,bib0014,bib0015,bib0016,bib0017,bib0018,bib0019,bib0020,bib0022}. Planck measurements strongly support this isotropic baseline while also motivating controlled searches for small, symmetry-classified departures \cite{bib0023,bib0024,bib0025}.

The lowest even-parity departure associated with one preferred direction is a tracefree quadrupole,
\begin{equation}
 \mathcal P_\zeta(\mathbf k)
 =\mathcal P_0(k)
 \left[1+g_*(k)\left((\hat{\mathbf k}\!\cdot\!\hat{\mathbf n})^2-\frac13\right)\right].
 \label{eq:main_quad_template}
\end{equation}
The subtraction of $1/3$ removes the isotropic shift in the scalar power and leaves only the spin-two angular part. Thus, Eq.~\eqref{eq:main_quad_template} is the natural observable for a model in which a covariant tensor develops a direction-dependent expectation value. In harmonic space, this structure produces off-diagonal CMB covariance and is naturally described with bipolar spherical harmonics (BipoSH) \cite{bib0031,bib0032,bib0033,bib0034}.

A complete theory must explain more than the angular template. It must identify the field that carries the preferred direction, state whether that field is dynamical or auxiliary, preserve the Hamiltonian and momentum constraints, and avoid negative-norm modes. It must also show when the anisotropy is generated, how its scale dependence follows from the background, and how the primordial signal reaches the CMB covariance. These checks are especially important for tensor fields: a covariant rank-two field is not automatically a healthy Fierz--Pauli spin-two particle. The standard massive spin-two constraint problem and a recent curvature-completed infinite-derivative realization show why the physical field space, constraints, kinetic operator, and residue signs must be stated explicitly \cite{bib0037,bib0038,bib0064}.

Curvature history provides the mechanism. The scalar $C$ records a covariant retarded response to the Gauss--Bonnet-completed tidal-curvature source
$\mathcal G_{\rm GB}=R_{\mu\nu\rho\sigma}R^{\mu\nu\rho\sigma}-4R_{\mu\nu}R^{\mu\nu}+R^2$. This source remains sensitive to the Kretschmann invariant in Ricci-flat regions and places the scalar--curvature coupling in the second-order Horndeski class. A conjugate response scalar $\chi$ gives a local variational form, while a symmetric memory tensor $\Delta_{\mu\nu}$ carries the state-dependent entire dressing. Relative to the timelike unit vector defined by the inflaton, $\Delta_{\mu\nu}$ splits into two scalars, a spatial vector, and a spatial symmetric tracefree tensor. Positive algebraic closure terms remove the non-tracefree parts, leaving
\begin{equation}
 Q_{\mu\nu}=\Pi_{\mu\nu}{}^{\alpha\beta}\Delta_{\alpha\beta},
 \qquad
 Q_{\mu\nu}u^\nu=0,
 \qquad
 Q^\mu{}_{\mu}=0,
 \label{eq:main_Q_definition_intro}
\end{equation}
as the physical tensor-memory order parameter. The five components of $Q_{\mu\nu}$ form a positive-norm spatial fiber; they are not the five helicities of a freely propagating massive spacetime spin-two field.

Curvature history controls a phase potential built from the independent invariants of a tracefree $3\times3$ tensor, in the standard spirit of invariant order-parameter expansions \cite{bib0035,bib0036}. On the activated branch the stable minimum is uniaxial,
\begin{equation}
 Q_{\mu\nu}^{\rm act}
 =D(C)\left(n_\mu n_\nu-\frac13h_{\mu\nu}\right),
 \qquad
 h_{\mu\nu}=g_{\mu\nu}+u_\mu u_\nu,
 \label{eq:main_uniaxial_intro}
\end{equation}
with $n_\mu u^\mu=0$ and $n_\mu n^\mu=1$. The preferred direction is therefore the nondegenerate eigenvector of a covariant order parameter. Unlike models that insert a spatial vector directly into the inflationary background, the direction here is selected by the tensor phase itself.

The memory phase affects scalar perturbations through a tracefree Ricci coupling. Diffeomorphism invariance fixes the energy--momentum exchange between the inflaton and memory--history sectors and conserves their total stress tensor. The ADM lapse and shift therefore remain constraint variables rather than new propagating modes. After these constraints are solved, the curvature perturbation acquires direction-dependent gradient energy. The anisotropic Mukhanov--Sasaki equation is stable only within a finite, exactly calculable range of the dimensionless memory amplitude. A finite activation period then produces an oscillatory window in $k$-space. Because the observable is the difference between the responses at the two endpoints, the constant infrared term cancels and $g_*(k)$ starts at order $k^2$.

\subsection{Relation to established mechanisms}\label{Xsec2-1.1}
\label{subsec:main_comparison}

Several established mechanisms produce the quadrupolar form in Eq.~\eqref{eq:main_quad_template}, but for different physical reasons. Gauge-field anisotropic inflation and the Ackerman--Carroll--Wise template use a vector background or vector hair \cite{bib0026,bib0027,bib0028}. Vector-curvaton models transfer anisotropy from a spectator vector sector \cite{bib0058,bib0059}. Solid inflation uses scalar fields with internal spatial symmetries, while primordial magnetic fields provide anisotropic electromagnetic stress \cite{bib0060,bib0061}. Generic BipoSH analyses can start directly from the correlator without specifying an action \cite{bib0062}. Here the direction is instead the principal axis of a component-closed tensor-memory phase sourced by curvature history. Table~\ref{tab:main_comparison} summarizes the comparison.

\begin{table}[t]
\centering
\caption{Structural comparison with common sources of primordial quadrupolar anisotropy. The distinction concerns the dynamical carrier and the origin of the observable kernel, not merely the angular template.}
\label{tab:main_comparison}
\begin{tabular}{llll}
\toprule
Mechanism & Direction carrier & Gravitational source & Scale dependence \\
\midrule
Gauge anisotropic inflation & vector hair & gauge stress & model/slow-roll dependent \\
Vector curvaton & spectator vector & transfer stress & transfer dependent \\
Solid inflation & scalar solid clocks & elastic stress & solid response \\
Primordial magnetic field & magnetic field & electromagnetic stress & magnetic template \\
Generic BipoSH & phenomenological $q_{LM}$ & unspecified & arbitrary $q_{LM}(k)$ \\
Tensor memory & eigenvector of $Q_{\mu\nu}$ & conserved memory--history stress & finite activation window \\
\botrule
\end{tabular}
\end{table}

The main result is a connected set of steps. These are covariant component closure, a dynamically selected activated branch, a projected spectral proof with no extra nonlocal poles, explicit ADM source closure, a stable anisotropic scalar gradient operator, an analytic infrared-regular window kernel, and a direct CMB BipoSH likelihood. Together, these steps derive the observable signal from the action rather than imposing a power-spectrum ansatz.

The paper proceeds from the covariant theory to its observable signal. Section~\ref{sec:main_action} defines the action and field content. Section~\ref{sec:main_eom} gives the field equations and conservation law. Section~\ref{sec:main_consistency} studies component closure, pole structure, and stability. Section~\ref{sec:main_background} treats homogeneous inflation, and Section~\ref{sec:main_perturbations} derives the activated phase and anisotropic scalar dynamics. Sections~\ref{sec:main_kernel}--\ref{sec:main_constraints} derive the primordial kernel, CMB observables, and likelihood. Section~\ref{sec:main_discriminants} compares the model with other mechanisms, while Secs.~\ref{sec:main_discussion} and \ref{sec:main_conclusion} discuss limitations and conclusions. Longer operator variations, background reductions, constraint calculations, and harmonic algebra are placed in Appendices~\ref{app:covariant}--\ref{app:biposh_likelihood}.

\section{Covariant action principle}\label{Xsec3-2}
\label{sec:main_action}

We work on a four-dimensional Lorentzian manifold $(\mathcal M,g_{\mu\nu})$ with signature $(-,+,+,+)$ and Levi--Civita connection $\nabla_\mu$ \cite{bib0006,bib0007}. The curvature conventions are
\begin{equation}
 [\nabla_\mu,\nabla_\nu]V^\rho=R^\rho{}_{\sigma\mu\nu}V^\sigma,
 \qquad
 G_{\mu\nu}=R_{\mu\nu}-\frac12g_{\mu\nu}R.
 \label{eq:main_curvature_conventions}
\end{equation}
The curvature-history source is chosen to be the four-dimensional Gauss--Bonnet completion of the Kretschmann invariant,
\begin{equation}
 \mathcal G_{\rm GB}
 =R_{\mu\nu\rho\sigma}R^{\mu\nu\rho\sigma}
 -4R_{\mu\nu}R^{\mu\nu}+R^2.
 \label{eq:main_GB_definition}
\end{equation}
With a constant coefficient, the integral is topological in four dimensions. With a scalar coefficient, it gives the scalar--Gauss--Bonnet member of the Horndeski class, so the metric and scalar equations contain at most second derivatives on a regular branch. In Ricci-flat geometries, $R_{\mu\nu}=R=0$, Eq.~\eqref{eq:main_GB_definition} reduces exactly to the Kretschmann scalar. On slow-roll FLRW, it differs from the Kretschmann source only at $O(\epsilon^2)$. This is the minimal curvature-squared completion that removes the nonlinear higher-derivative metric mode of an isolated $\chi R_{\mu\nu\rho\sigma}R^{\mu\nu\rho\sigma}$ coupling \cite{bib0004,bib0005}.

The field content is
\begin{equation}
 \{g_{\mu\nu},\phi,C,\chi,\Delta_{\mu\nu}\},
 \qquad
 \Delta_{\mu\nu}=\Delta_{\nu\mu},
 \label{eq:main_fields}
\end{equation}
where $\Delta_{\mu\nu}$ is a covariant parent tensor.  The inflaton gradient defines the physical time direction and the associated spatial projectors,
\begin{equation}
 u_\mu=-\frac{\nabla_\mu\phi}{\sqrt{-\nabla_\rho\phi\nabla^\rho\phi}},
 \qquad
 h_{\mu\nu}=g_{\mu\nu}+u_\mu u_\nu,
 \qquad
 \Pi_{\mu\nu}{}^{\alpha\beta}
 =h_{(\mu}{}^\alpha h_{\nu)}{}^\beta-\frac13h_{\mu\nu}h^{\alpha\beta}.
 \label{eq:main_projector}
\end{equation}
Relative to this direction,
\begin{equation}
 \Delta_{\mu\nu}=Au_\mu u_\nu+2u_{(\mu}B_{\nu)}
 +\frac13\Theta h_{\mu\nu}+Q_{\mu\nu},
 \qquad
 \widehat Q_{\mu\nu}=\Pi_{\mu\nu}{}^{\alpha\beta}\Delta_{\alpha\beta},
 \label{eq:main_decomposition}
\end{equation}
with $B_\mu u^\mu=0$, $\widehat Q_{\mu\nu}u^\nu=0$, and $\widehat Q^\mu{}_\mu=0$.  This decomposition separates the five raw spatial tracefree components from the five complementary components.  The dressed physical order parameter is defined in Eq.~\eqref{eq:main_dressed_Q}.

The total action is
\begin{equation}
 S=S_{\rm EH}+S_\phi+S_C+S_\Delta.
 \label{eq:main_total_action}
\end{equation}
The Einstein--Hilbert and inflaton sectors take their standard forms,
\begin{align}
 S_{\rm EH}&=\frac{M_{\rm Pl}^2}{2}\int d^4x\sqrt{-g}\,R,\\
 S_\phi&=\int d^4x\sqrt{-g}\left[-\frac12(\nabla\phi)^2-U(\phi)\right].
 \label{eq:main_standard_actions}
\end{align}
The local response representation is
\begin{equation}
 S_C=\int d^4x\sqrt{-g}\left[-\nabla_\mu\chi\nabla^\mu C-m_c^2\chi C+\ell_*\chi\mathcal G_{\rm GB}\right].
 \label{eq:main_SC}
\end{equation}
If treated as an unconstrained local two-scalar theory, this action has an indefinite kinetic matrix. The physical theory instead uses the retarded prescription in Section~\ref{subsec:retarded_phase_space}: $C$ and $\chi$ have fixed homogeneous data and act as response variables, not independently quantized asymptotic fields. Calling the fields auxiliary does not by itself impose this restriction \cite{bib0002,bib0003}.

The raw STF projection is
\begin{equation}
 \widehat Q_{\mu\nu}=\Pi_{\mu\nu}{}^{\alpha\beta}\Delta_{\alpha\beta},
 \qquad
 (\Box_{\widehat Q}X)_{\mu\nu}
 =\Pi_{\mu\nu}{}^{\alpha\beta}g^{\rho\sigma}\nabla_\rho\nabla_\sigma
 (\Pi_{\alpha\beta}{}^{\gamma\delta}X_{\gamma\delta}).
 \label{eq:main_raw_Q}
\end{equation}
Define the ordered half-form-factor and the dressed physical field by
\begin{equation}
 \mathcal E_C
 =\exp\!\left[\frac12 B(C)\Box_{\widehat Q}\right],
 \qquad
 B(C)=M_{\rm eff}^{-2}(C)>0,
 \qquad
 Q_{\mu\nu}=\mathcal E_C\widehat Q_{\mu\nu}.
 \label{eq:main_dressed_Q}
\end{equation}
The exponential is invertible on the common functional domain $\mathscr D$, with inverse $\mathcal E_C^{-1}=\exp[-B(C)\Box_{\widehat Q}/2]$ defined by the same ordered power series.  Because the projected operator maps the STF bundle into itself, $Q_{\mu\nu}$ is again spatial and tracefree.

The component-closed memory action is written in the dressed variable,
\begin{align}
 S_\Delta=\int d^4x\sqrt{-g}\Bigl[&
 +\frac12Q_{\mu\nu}(\Box_Q Q)^{\mu\nu}
 -V_Q(Q,C)
 -\frac12M_A^2A^2-\frac12M_B^2B_\mu B^\mu-\frac12M_\Theta^2\Theta^2
 \notag\\
 &+\eta G(C)Q_{\mu\nu}R^{\mu\nu}\Bigr],
 \label{eq:main_SDelta}
\end{align}
where $\Box_Q$ denotes the formally self-adjoint projected d'Alembertian acting on the dressed STF field.  The local kinetic term has the healthy $(-,+,+,+)$ sign: up to a boundary term,
\begin{equation}
 \frac12\langle Q,\Box_QQ\rangle
 =-\frac12\langle\nabla Q,\nabla Q\rangle,
 \label{eq:main_healthy_sign}
\end{equation}
so its local orthonormal-frame kinetic energy is $+\tfrac12\sum_A\dot q_A^2$. In terms of the raw field $\widehat Q$, the Hessian is the self-adjoint sandwich $\mathcal E_C^\dagger(\Box_Q-\mathsf M^2)\mathcal E_C$. This factorization follows directly from the action and gives the corresponding propagator.

Because $B(C)$ and $\Box_{\widehat Q}$ do not commute when $C$ varies in spacetime, the operator order in Eq.~\eqref{eq:main_dressed_Q} is part of the definition. We vary it with Duhamel's formula using the ordered-operator treatment of state-dependent nonlocality \cite{bib0001,bib0046,bib0047}. The adjoint and the time-dependent spectral analysis are given in Section~\ref{subsec:time_dependent_factorization} and Appendix~\ref{app:closure}.

\paragraph*{Ordered state-dependent dressing}

Put
\begin{equation}
 \mathcal A_C\equiv\frac12B(C)\Box_{\widehat Q},
 \qquad
 \mathcal E_C=e^{\mathcal A_C}
 =\sum_{n=0}^{\infty}\frac{\mathcal A_C^n}{n!}.
 \label{eq:ordered_half_factor}
\end{equation}
The product acts from right to left.  Since multiplication by $B(C)$ does not commute with the projected differential operator,
\begin{equation}
 [\Box_{\widehat Q},B]X_{\mu\nu}
 =\Pi_{\mu\nu}{}^{\alpha\beta}
 \left[(\Box B)X_{\alpha\beta}
 +2(\nabla^\rho B)\nabla_\rho X_{\alpha\beta}\right]
 +\mathcal C_{\mu\nu}[\nabla\Pi,B,X],
 \label{eq:boxB_commutator}
\end{equation}
where $\mathcal C_{\mu\nu}$ contains the projector-derivative terms.  For any variation of the background data the exact Fr\'echet derivative is the Duhamel formula
\begin{equation}
 \delta\mathcal E_C
 =\int_0^1ds\,
 e^{s\mathcal A_C}(\delta\mathcal A_C)e^{(1-s)\mathcal A_C},
 \qquad
 \delta\mathcal A_C
 =\frac12B_{,C}\,\delta C\,\Box_{\widehat Q}
 +\frac12B\,\delta\Box_{\widehat Q}.
 \label{eq:Duhamel_E}
\end{equation}
With the tensor pairing $\langle X,Y\rangle=\int\sqrt{-g}\,X_{\mu\nu}Y^{\mu\nu}$,
\begin{equation}
 \mathcal A_C^\dagger=\frac12\Box_{\widehat Q}^{\dagger}B(C),
 \qquad
 \mathcal E_C^\dagger=e^{\mathcal A_C^\dagger}.
 \label{eq:adjoint_E}
\end{equation}
These identities are the state-dependent auxiliary-field/operator mechanism used in the covariant variation; the additional retarded restriction in Section~\ref{subsec:retarded_phase_space} is what reduces the localized response phase space.

The coupling $\eta G(C)Q_{\mu\nu}R^{\mu\nu}$ has two roles. It lets curvature history affect the preferred tensor phase, and once a tracefree expectation value forms, it changes the scalar gradient sector without adding a monopole term. Because the coupling is covariant, its ADM contributions follow from varying the full action with respect to lapse and shift. No separate anisotropic source is added by hand.

We assume that $B(C)$ and $G(C)$ are real and smooth, the closure masses are positive on the background of interest, and the Hessian of $V_Q$ is positive on the selected activated branch. These conditions separate three different issues: algebraic removal of auxiliary components, the pole structure of the projected kinetic operator, and local stability of the phase potential. Section~\ref{sec:main_consistency} summarizes the results, and Appendix~\ref{app:closure} gives the detailed proof.

\subsection{Retarded physical phase space for \texorpdfstring{$(C,\chi)$}{(c, chi)}}\label{Xsec4-2.1}
\label{subsec:retarded_phase_space}

The local action \eqref{eq:main_SC} is a variational localization.  It is not the definition of an unconstrained two-scalar Hilbert space.  Put
\begin{equation}
 L_c\equiv\Box-m_c^2,
 \qquad
 L_cG_{\rm ret}(x,x')=\frac{\delta^{(4)}(x-x')}{\sqrt{-g(x')}} ,
 \qquad
 \operatorname{supp}G_{\rm ret}(x,\cdot)\subset J^-(x).
 \label{eq:retarded_green}
\end{equation}
For fixed homogeneous data on an initial Cauchy surface $\Sigma_*$, define the physical response map
\begin{align}
 C[g](x)&=C_{\rm hom}^{\rm fixed}(x)
 -\ell_*\!\int d^4x'\sqrt{-g'}\,G_{\rm ret}(x,x')\mathcal G_{\rm GB}(x'),
 \label{eq:C_retarded_solution}\\
 \chi[g,Q](x)&=\chi_{\rm hom}^{\rm fixed}(x)
 -\!\int d^4x'\sqrt{-g'}\,G_{\rm ret}(x,x')\Sigma_\Delta(x').
 \label{eq:chi_retarded_solution}
\end{align}
The physical phase space is therefore the graph
\begin{equation}
 \Gamma_{\rm phys}
 =\left\{(g,\phi,Q;C[g],\chi[g,Q])\right\}
 \subset\Gamma_{\rm loc},
 \label{eq:physical_graph_phase_space}
\end{equation}
not the full localized phase space $\Gamma_{\rm loc}$.  Tangent vectors obey
\begin{equation}
 c|_{\Sigma_*}=n^\mu\nabla_\mu c|_{\Sigma_*}
 =x|_{\Sigma_*}=n^\mu\nabla_\mu x|_{\Sigma_*}=0,
 \qquad c\equiv\delta C,\quad x\equiv\delta\chi.
 \label{eq:retarded_initial_constraints}
\end{equation}
These are boundary restrictions defining the retarded theory; they are neither gauge conditions nor Dirac constraints of the larger localized theory \cite{bib0002,bib0003}.

The unrestricted localized kinetic term indeed has indefinite signature:
\begin{equation}
 a^{-3}L_{C\chi}^{\rm kin}=\dot\chi\dot C
 =\frac12\dot\psi^2-\frac12\dot\xi^2,
 \qquad
 \psi=\frac{C+\chi}{\sqrt2},\qquad
 \xi=\frac{C-\chi}{\sqrt2}.
 \label{eq:local_indefinite_pair}
\end{equation}
For an unconstrained localization, the kinetic sector therefore contains one positive-sign and one negative-sign combination. The localized canonical momenta and symplectic form are
\begin{equation}
 \pi_C=a^3\dot\chi,
 \qquad
 \pi_\chi=a^3\dot C,
 \qquad
 \Omega_{C\chi}=\int_{\Sigma}d^3x\,
 \bigl(\delta\pi_C\wedge\delta C+\delta\pi_\chi\wedge\delta\chi\bigr).
 \label{eq:aux_symplectic}
\end{equation}

\begin{myproposition}[Elimination of the localized ghost direction]
\label{prop:retarded_reduction}
Let $d\iota_{\rm ret}:T\Gamma_{\rm phys}\rightarrow T\Gamma_{\rm loc}$ be the tangent map of the embedding \eqref{eq:physical_graph_phase_space}. If the physical variations vanish, $\delta g=\delta\phi=\delta Q=0$, then $c=x=0$. Hence the two homogeneous response directions, including the negative-sign $\xi$ direction in \eqref{eq:local_indefinite_pair}, do not define tangent directions of $T\Gamma_{\rm phys}$.
\label{Xenun1-1}
\end{myproposition}

\begin{proof}
If $\delta g=\delta\phi=\delta Q=0$, the linearized sources in \eqref{eq:C_retarded_solution} and \eqref{eq:chi_retarded_solution} vanish, so $L_cc=0=L_cx$. Eq.~\eqref{eq:retarded_initial_constraints} also sets the field values and normal derivatives to zero. Uniqueness of the normally hyperbolic Cauchy problem then gives $c=x=0$. Thus no nonzero homogeneous $\psi$ or $\xi$ solution is tangent to the physical graph. In particular, the negative-sign localized direction has no independent canonical coordinate, momentum, creation operator, or asymptotic state.
\end{proof}

The physical symplectic form is the pullback
\begin{equation}
 \Omega_{\rm phys}=\iota_{\rm ret}^{*}\Omega_{\rm loc}
 =\Omega_{g\phi Q}+\iota_{\rm ret}^{*}\Omega_{C\chi}.
 \label{eq:physical_symplectic_pullback}
\end{equation}
The second term can renormalize the symplectic form because $c$ and $x$ are induced by the propagating variables. However, Proposition~\ref{prop:retarded_reduction} shows that it cannot add a new positive- or negative-norm response state. In the benchmark solution, the response energy fraction is below $1.2\times10^{-7}$, so this correction is perturbative and the inflaton and STF kinetic blocks remain strictly positive. The causal equations are interpreted in the retarded/in-in sense \cite{bib0048,bib0049}. Substituting \eqref{eq:C_retarded_solution} and \eqref{eq:chi_retarded_solution} into an ordinary in-out action would not define the retarded theory.

\section{Variational dynamics and field equations}\label{Xsec5-3}
\label{sec:main_eom}

The equations of motion follow by independent variation with respect to $g_{\mu\nu}$, $\phi$, $C$, $\chi$, and $\Delta_{\mu\nu}$. The detailed integrations by parts and the noncommuting form-factor variation are given in Appendix~\ref{app:covariant}. These equations control the physical flow of the theory.

Variation with respect to $\chi$ gives the curvature-history equation
\begin{equation}
 (\Box-m_c^2)C=-\ell_*\mathcal G_{\rm GB}.
 \label{eq:main_C_eom}
\end{equation}
This equation gives a direct meaning to the history field: it is a retarded massive response to the Gauss--Bonnet-completed tidal-curvature invariant. In Ricci-flat regions, the source is exactly the Kretschmann scalar. The mass $m_c$ sets the spacetime memory range, and $\ell_*$ sets the conversion from curvature squared to the history amplitude. Boundary or initial data select the Green-function solution, while the source is fixed by the geometry.

Variation with respect to $C$ at fixed raw field $\widehat Q$ gives
\begin{equation}
 (\Box-m_c^2)\chi=-\Sigma_\Delta,
 \label{eq:main_chi_eom}
\end{equation}
where the exact source is defined by
\begin{align}
 \Sigma_\Delta
 &\equiv\frac{1}{\sqrt{-g}}\left.\frac{\delta S_\Delta}{\delta C}\right|_{\widehat Q,g,\phi}
 \notag\\
 &=-V_{Q,C}
 -\frac12(M_A^2)_{,C}A^2
 -\frac12(M_B^2)_{,C}B_\mu B^\mu
 -\frac12(M_\Theta^2)_{,C}\Theta^2
 +\eta G_{,C}Q_{\mu\nu}R^{\mu\nu}
 +\Sigma_{\rm dress}.
 \label{eq:main_Sigma}
\end{align}
The dressing contribution is most compactly and exactly specified by
\begin{equation}
 \int d^4x\sqrt{-g}\,\Sigma_{\rm dress}\,\delta C
 =\left\langle\mathcal E^{(Q)},(\delta_C\mathcal E_C)\widehat Q\right\rangle,
 \qquad
 \mathcal E^{(Q)}_{\mu\nu}
 =\Box_QQ_{\mu\nu}-V_{Q,Q^{\mu\nu}}
 +\eta G\Pi_{\mu\nu}{}^{\alpha\beta}R_{\alpha\beta},
 \label{eq:Sigma_dress_exact}
\end{equation}
with $\delta_C\mathcal E_C$ given by Eq.~\eqref{eq:Duhamel_E}.  Thus $\Sigma_{\rm dress}$ vanishes on the dressed $Q$ equation of motion, while its off-shell form retains all noncommuting state-dependent terms.

Variation with respect to the raw projected field uses $\delta Q=\mathcal E_C\delta\widehat Q$ at fixed $(g,\phi,C)$.  Hence
\begin{equation}
 \mathcal E_C^\dagger\mathcal E_{\mu\nu}^{(Q)}=0,
 \qquad
 \mathcal E_{\mu\nu}^{(Q)}
 =\Box_QQ_{\mu\nu}-\frac{\partial V_Q}{\partial Q^{\mu\nu}}
 +\eta G(C)\Pi_{\mu\nu}{}^{\alpha\beta}R_{\alpha\beta}.
 \label{eq:main_Delta_eom}
\end{equation}
Because $\mathcal E_C^\dagger$ is invertible on $\mathscr D$, Eq.~\eqref{eq:main_Delta_eom} is equivalent to the second-order dressed-field equation $\mathcal E_{\mu\nu}^{(Q)}=0$. The state-dependent entire dressing therefore changes the raw-field propagator and ultraviolet weighting without adding another solution branch, as in the zero-free form-factor approach to infinite-derivative theories \cite{bib0040,bib0041,bib0042,bib0043,bib0044,bib0045}. The complementary equations remain algebraic,
\begin{equation}
 M_A^2A=0,\qquad M_B^2B_\mu=0,\qquad M_\Theta^2\Theta=0,
 \label{eq:main_closure_field_equations}
\end{equation}
so positive closure masses uniquely select $A=B_\mu=\Theta=0$.

The inflaton satisfies
\begin{equation}
 \Box\phi-U_{,\phi}=\mathcal J_\phi^{(\Delta)},
 \label{eq:main_phi_eom}
\end{equation}
where $\mathcal J_\phi^{(\Delta)}$ comes from the dependence of the decomposition projectors on the inflaton-defined normal. On a homogeneous, component-closed branch, it reduces consistently to the background stress-energy exchange. Keeping this term covariant avoids mixing a covariant tensor action with a preferred frame imposed separately.

Metric variation gives the Einstein equation
\begin{equation}
 M_{\rm Pl}^2G_{\mu\nu}=T_{\mu\nu}^{(\phi)}+T_{\mu\nu}^{(C)}+T_{\mu\nu}^{(\Delta)}.
 \label{eq:main_Einstein}
\end{equation}
The individual history and memory tensors need not be conserved separately because $C$ enters the memory operator and the coupling functions. Diffeomorphism invariance instead gives the on-shell exchange identity
\begin{equation}
 \nabla^\mu\left(T_{\mu\nu}^{(C)}+T_{\mu\nu}^{(\Delta)}\right)
 =-\mathcal J_\phi^{(\Delta)}\nabla_\nu\phi,
 \label{eq:main_combined_conservation}
\end{equation}
while the inflaton tensor obeys $\nabla^\mu T_{\mu\nu}^{(\phi)}=\mathcal J_\phi^{(\Delta)}\nabla_\nu\phi$. Their sum is exactly conserved. This exchange law supplies the source for both the covariant Einstein equation and the ADM constraints. Treating $T_{\mu\nu}^{(\Delta)}$ as conserved by itself would miss energy--momentum exchanged through the curvature-history dependence and the inflaton-defined projector.

Projecting Eq.~\eqref{eq:main_Einstein} along the unit normal gives the Hamiltonian and momentum constraints,
\begin{align}
 \frac{M_{\rm Pl}^2}{2}\left({}^{(3)}R+K^2-K_{ij}K^{ij}\right)&=\mathcal E_\phi+\mathcal E_{\rm mem},\\
 M_{\rm Pl}^2D_j\left(K^j{}_i-\delta^j_iK\right)&=\mathcal J_{i,\phi}+\mathcal J_{i,{\rm mem}},
 \label{eq:main_ADM_constraints_cov}
\end{align}
where ``mem'' denotes the combined $C+\Delta$ sector. These equations retain the standard geometric operators. New physics appears through covariantly derived sources, not through a modification of the constraint algebra.

The variational equations have a clear order. Eq.~\eqref{eq:main_C_eom} determines the curvature-history background, Eq.~\eqref{eq:main_Delta_eom} determines the tensor-phase response, and Eq.~\eqref{eq:main_combined_conservation} ensures consistent gravitational sourcing. The ADM constraints in Eqs.~\eqref{eq:main_ADM_constraints_cov} then remove the nondynamical metric variables before the scalar action is reduced. This order is important for the physical interpretation of the model.

\section{Theoretical consistency and stability}\label{Xsec6-4}
\label{sec:main_consistency}

A viable tensor-memory theory must pass three separate consistency tests. The covariant symmetric tensor must not leave unaccounted components; the nonlocal form factor must not introduce additional ghost poles; and the activated background must have positive local fluctuation energies. Component closure, pole structure, and local stability are examined in that order.

From the decomposition \eqref{eq:main_decomposition}, the covariant tensor contains $1+3+1+5=10$ components. The closure terms in Eq.~\eqref{eq:main_SDelta} supply algebraic equations
\begin{equation}
 M_A^2A=0,
 \qquad
 M_B^2B_\mu=0,
 \qquad
 M_\Theta^2\Theta=0.
 \label{eq:main_closure_eom}
\end{equation}
For positive closure functions, the unique regular branch is
\begin{equation}
 A=0,
 \qquad
 B_\mu=0,
 \qquad
 \Theta=0.
 \label{eq:main_aux_elimination}
\end{equation}
The remaining configuration space is the five-dimensional spatial symmetric tracefree bundle. Its natural norm is positive because the induced metric $h_{\mu\nu}$ is positive definite. This removes the ambiguity that would arise if the temporal and longitudinal components of an unconstrained spacetime tensor were treated as physical.

Let $Q_{\mu\nu}=\bar Q_{\mu\nu}+q_{\mu\nu}$ be expanded about a stationary dressed background and let $\mathsf M^2$ be the STF Hessian of $V_Q$.  In raw variables $\widehat q$, the exact quadratic operator is
\begin{equation}
 \mathcal K_{\rm raw}
 =\mathcal E_C^\dagger(\Box_Q-\mathsf M^2)\mathcal E_C.
 \label{eq:main_Kproj}
\end{equation}
On a constant-history background, $C=C_0$, the operators commute and $\mathcal E_C^\dagger=\mathcal E_C=e^{B_0\Box_Q/2}$.  In an STF Hessian eigenbasis this gives
\begin{equation}
 \mathcal K_A(z)=e^{B_0z}(z-m_A^2),
 \qquad
 \mathcal G_A(z)=\frac{e^{-B_0z}}{z-m_A^2},
 \label{eq:main_constant_propagator}
\end{equation}
so the unique pole is $z=m_A^2$ and
\begin{equation}
 {\rm Res}_{z=m_A^2}\mathcal G_A=e^{-B_0m_A^2}>0.
 \label{eq:main_residue}
\end{equation}
The same no-extra-branch statement in fact survives arbitrary smooth $C(x)$ because it follows from invertibility of the sandwich factors rather than from commutativity; Section~\ref{subsec:time_dependent_factorization} gives the exact argument and its WKB interpretation.

The explicit Hessian decomposition and spectral argument identify the fluctuation masses and show where the zero-free property removes extra poles:

\paragraph*{Quadratic Hessian of the activated phase}

The invariant phase potential for the tracefree order parameter is
\begin{equation}
  V_Q(Q,C)
  =
  V_0(C)
  +
  \frac12\mu^2(C)I_2
  -
  \frac13\gamma(C)I_3
  +
  \frac14\lambda(C)I_2^2,
  \label{msp:eq:VQ_section5}
\end{equation}
with
\begin{equation}
  I_2=Q_{\mu\nu}Q^{\mu\nu},
  \qquad
  I_3=Q_\mu{}^\nu Q_\nu{}^\rho Q_\rho{}^\mu .
  \label{msp:eq:I2I3_section5}
\end{equation}
The first derivative in the tracefree spatial subspace is
\begin{equation}
  \mathcal{F}_{\mu\nu}
  =
  \left[
    \mu^2+\lambda I_2
  \right]Q_{\mu\nu}
  -
  \gamma
  \left(
    Q_\mu{}^\rho Q_{\rho\nu}
    -
    \frac13I_2h_{\mu\nu}
  \right).
  \label{msp:eq:F_phase_section5}
\end{equation}
The activated background obeys
\begin{equation}
  \mathcal{F}_{\mu\nu}[Q^{(0)},C_0]=0 .
  \label{msp:eq:F_zero_Q0_section5}
\end{equation}

The Hessian is the linear map $\mathsf{M}^2:\mathscr{S}_0^2\rightarrow\mathscr{S}_0^2$ defined by
\begin{equation}
  (\mathsf{M}^2q)_{\mu\nu}
  =
  \left.
  \frac{d}{d\varepsilon}
  \mathcal{F}_{\mu\nu}
  [Q^{(0)}+\varepsilon q,C_0]
  \right|_{\varepsilon=0}.
  \label{msp:eq:Hessian_definition_section5}
\end{equation}
Using
\begin{equation}
  \delta I_2=2Q^{\alpha\beta}q_{\alpha\beta},
\label{Xeqn42-50}
\end{equation}
one obtains
\begin{align}
  (\mathsf{M}^2q)_{\mu\nu}
  &=
  \left[
    \mu_0^2+\lambda_0 I_2^{(0)}
  \right]q_{\mu\nu}
  +
  2\lambda_0
  \left(
    Q^{(0)\alpha\beta}q_{\alpha\beta}
  \right)
  Q_{\mu\nu}^{(0)}
  \notag\\
  &\quad
  -
  \gamma_0
  \left[
    Q_\mu^{(0)\rho}q_{\rho\nu}
    +
    q_\mu{}^\rho Q_{\rho\nu}^{(0)}
    -
    \frac23
    \left(
      Q^{(0)\alpha\beta}q_{\alpha\beta}
    \right)
    h_{\mu\nu}
  \right],
  \label{msp:eq:Hessian_operator_section5}
\end{align}
where
\begin{equation}
  I_2^{(0)}
  =
  Q_{\mu\nu}^{(0)}Q^{(0)\mu\nu}
  =
  \frac23D_0^2 .
  \label{msp:eq:I20_section5}
\end{equation}
Using Eq.~\eqref{eq:stationary_D0_section5},
\begin{equation}
  \mu_0^2+\lambda_0I_2^{(0)}
  =
  \mu_0^2+\frac23\lambda_0D_0^2
  =
  \frac13\gamma_0D_0 .
  \label{msp:eq:stationary_simplification_section5}
\end{equation}

Choose an orthonormal spatial triad
\begin{equation}
  \{n^\mu,e_1^\mu,e_2^\mu\},
  \qquad
  e_A^\mu u_\mu=0,
  \qquad
  e_A^\mu n_\mu=0,
  \qquad
  e_A^\mu e_{B\mu}=\delta_{AB}.
\label{Xeqn45-54}
\end{equation}
A normalized basis of $\mathscr{S}_0^2$ is
\begin{align}
  E_0{}_{\mu\nu}
  &=
  \sqrt{\frac32}
  \left(
    n_\mu n_\nu
    -
    \frac13h_{\mu\nu}
  \right),
  \label{msp:eq:E0_section5}
  \\
  E_1{}_{\mu\nu}
  &=
  \sqrt{2}\,n_{(\mu}e_{1\nu)},
  \qquad
  E_2{}_{\mu\nu}
  =
  \sqrt{2}\,n_{(\mu}e_{2\nu)},
  \label{msp:eq:E12_section5}
  \\
  E_3{}_{\mu\nu}
  &=
  \frac{1}{\sqrt{2}}
  \left(
    e_{1\mu}e_{1\nu}
    -
    e_{2\mu}e_{2\nu}
  \right),
  \label{msp:eq:E3_section5}
  \\
  E_4{}_{\mu\nu}
  &=
  \sqrt{2}\,e_{1(\mu}e_{2\nu)}.
  \label{msp:eq:E4_section5}
\end{align}
These tensors obey
\begin{equation}
  E_A{}_{\mu\nu}E_B{}^{\mu\nu}=\delta_{AB},
  \qquad
  A,B=0,\ldots,4.
  \label{msp:eq:E_orthonormal_section5}
\end{equation}
Expanding
\begin{equation}
  q_{\mu\nu}
  =
  \sum_{A=0}^{4}q_AE_A{}_{\mu\nu},
  \label{msp:eq:q_basis_expansion_section5}
\end{equation}
the Hessian is diagonal:
\begin{equation}
  \mathsf{M}^2E_A
  =
  m_A^2E_A .
  \label{msp:eq:Hessian_diagonal_section5}
\end{equation}

The eigenvalues are obtained by substituting each basis tensor into Eq.~\eqref{msp:eq:Hessian_operator_section5}. The amplitude mode $E_0$ gives
\begin{equation}
  m_0^2
  =
  \frac{D_0}{3}
  \left(
    4\lambda_0D_0-\gamma_0
  \right).
  \label{msp:eq:m0_section5}
\end{equation}
Using Eq.~\eqref{eq:D0_section5},
\begin{equation}
  4\lambda_0D_0-\gamma_0
  =
  \sqrt{\gamma_0^2-24\lambda_0\mu_0^2},
\label{Xeqn50-63}
\end{equation}
so
\begin{equation}
  m_0^2
  =
  \frac{D_0}{3}
  \sqrt{\gamma_0^2-24\lambda_0\mu_0^2}
  >
  0 .
  \label{msp:eq:m0_positive_section5}
\end{equation}
The two orientational modes $E_1$ and $E_2$ give
\begin{equation}
  m_1^2=m_2^2=0 .
  \label{msp:eq:orientational_masses_section5}
\end{equation}
These are the two Goldstone directions associated with
\begin{equation}
  SO(3)\rightarrow SO(2).
\label{Xeqn53-66}
\end{equation}
The two biaxial modes $E_3$ and $E_4$ give
\begin{equation}
  m_3^2=m_4^2=\gamma_0D_0>0 .
  \label{msp:eq:biaxial_masses_section5}
\end{equation}

Thus the projected tensor-memory spectrum contains one massive amplitude mode, two massless orientational modes, and two massive biaxial modes:
\begin{equation}
  \{m_A^2\}_{A=0}^{4}
  =
  \left\{
    \frac{D_0}{3}
    \sqrt{\gamma_0^2-24\lambda_0\mu_0^2},
    0,
    0,
    \gamma_0D_0,
    \gamma_0D_0
  \right\}.
  \label{msp:eq:mass_spectrum_section5}
\end{equation}

\paragraph*{Quadratic sign and exact factorization}

In the orthonormal STF Hessian basis, let $\widehat q_A$ denote the raw fluctuation and $q_A=\mathcal E_C\widehat q_A$ the dressed fluctuation.  The quadratic action is
\begin{equation}
 S_q^{(2)}
 =+\frac12\sum_{A=0}^{4}\int d^4x\sqrt{-g}\,
 \widehat q_A\,
 \mathcal E_C^\dagger(\Box_Q-m_A^2)\mathcal E_C\,
 \widehat q_A.
 \label{msp:eq:quadratic_action_ghostfree_section5}
\end{equation}
For $B\to0$ this becomes
\begin{align}
 S_q^{(2)}&=+\frac12\sum_A\int\sqrt{-g}\,q_A(\Box_Q-m_A^2)q_A
 \notag\\
 &\simeq\sum_A\int\sqrt{-g}\left[-\frac12\nabla_\mu q_A\nabla^\mu q_A-\frac12m_A^2q_A^2\right]
 \notag\\
 &=\sum_A\int dt\,d^3x\,a^3
 \left[+\frac12\dot q_A^2-\frac{1}{2a^2}(\partial_iq_A)^2-\frac12m_A^2q_A^2\right],
 \label{msp:eq:healthy_local_limit_section5}
\end{align}
which has the healthy local sign for the $(-,+,+,+)$ convention.

On $C=C_0$, Eq.~\eqref{msp:eq:quadratic_action_ghostfree_section5} reduces to
\begin{equation}
 S_q^{(2)}=+\frac12\sum_A\int\sqrt{-g}\,
 \widehat q_Ae^{B_0\Box_Q}(\Box_Q-m_A^2)\widehat q_A,
 \label{msp:eq:constant_factorized_action_section5}
\end{equation}
with propagator $e^{-B_0z}/(z-m_A^2)$ and positive residue. The factorized mass term follows directly from the dressed-field action rather than being imposed after expanding the potential.

\subsection{Time-dependent factorization, symplectic norm, and WKB poles}\label{Xsec7-4.1}
\label{subsec:time_dependent_factorization}

The constant-history pole argument is not needed to count modes.  For arbitrary smooth $C(x)$ define
\begin{equation}
 L_A\equiv\Box_Q-m_A^2(x),
 \qquad E\equiv\mathcal E_C,
 \qquad K_A=E^\dagger L_AE.
 \label{eq:time_dependent_sandwich}
\end{equation}
We define the entire-function theory on the common invariant space $\mathscr D_{\rm an}$ of analytic vectors of $\mathcal A_C=B(C)\Box_{\widehat Q}/2$ and $\mathcal A_C^\dagger$, stable under $L_A$.  On this domain the exponential series and its inverse converge and
\begin{equation}
 E^{-1}=e^{-\mathcal A_C},\qquad (E^\dagger)^{-1}=e^{-\mathcal A_C^\dagger},
 \qquad E^{-1}E=EE^{-1}=1.
 \label{eq:exact_E_inverse}
\end{equation}
Consequently
\begin{equation}
 K_A\widehat q_A=0
 \quad\Longleftrightarrow\quad
 L_Aq_A=0,
 \qquad q_A=E\widehat q_A,
 \label{eq:kernel_bijection}
\end{equation}
and therefore
\begin{equation}
 \ker K_A=E^{-1}(\ker L_A),
 \qquad
 \dim\ker K_A=\dim\ker L_A.
 \label{eq:kernel_dimension}
\end{equation}
This is an exact statement and does not use $[B(C),\Box_Q]=0$.

The inverse kernel has the exact congruence form
\begin{equation}
 G_K=E^{-1}G_L(E^\dagger)^{-1},
 \qquad L_AG_L=1,
 \label{eq:green_congruence}
\end{equation}
with the same physical contour/boundary prescription imposed in the dressed variable.  The covariant symplectic form of the local dressed equation is
\begin{equation}
 \Omega_L(q_1,q_2)=\int_\Sigma d\Sigma_\mu
 \left(q_1\nabla^\mu q_2-q_2\nabla^\mu q_1\right).
 \label{eq:local_symplectic_q}
\end{equation}
Define the raw-field norm by pullback,
\begin{equation}
 \Omega_K(\widehat q_1,\widehat q_2)
 \equiv\Omega_L(E\widehat q_1,E\widehat q_2).
 \label{eq:symplectic_congruence}
\end{equation}
Because $E$ is bijective, \eqref{eq:symplectic_congruence} is a congruence transformation and preserves the inertia of the mode norm.  Hence every positive-norm dressed solution maps to a positive-norm raw solution and no negative-norm branch is generated by a time-dependent $B(C)$.

A local quasiparticle-pole description follows from the Weyl symbol on a local orthonormal patch. Let $z=\omega^2-\mathbf p^2$, $e(t,z)=\exp[B(t)z/2]$, and $\ell_A(t,z)=z-M_A(t)$ with $M_A(t)=m_A^2(t)$. The Moyal product gives
\begin{equation}
 k_A=e\star\ell_A\star e
 =e^{Bz}(z-M_A)+\Delta_A^{(2)}+O(\partial_t^3),
 \label{eq:wkb_symbol_second_order}
\end{equation}
where the complete first adiabatic order cancels identically.  Writing
\begin{equation}
 \Lambda(a,b)=a_{,t}b_{,\omega}-a_{,\omega}b_{,t},
 \qquad
 \Lambda^2(a,b)=a_{,tt}b_{,\omega\omega}
 -2a_{,t\omega}b_{,t\omega}+a_{,\omega\omega}b_{,tt},
 \label{eq:moyal_bidifferential}
\end{equation}
the exact second-order term is
\begin{equation}
 \Delta_A^{(2)}=-\frac18 e\,\Lambda^2(e,\ell_A)
 -\frac14\Lambda\!\left(\Lambda(e,\ell_A),e\right)
 -\frac18\Lambda^2(e\ell_A,e).
 \label{eq:Delta2_exact}
\end{equation}
The quasiparticle zero and residue are therefore
\begin{align}
 z_A^\star&=M_A-e^{-BM_A}\Delta_A^{(2)}\big|_{z=M_A}+O(\partial_t^3),
 \label{eq:wkb_pole_shift}\\
 Z_A&=e^{-BM_A}\left[1-\delta_A^{(2)}+O(\partial_t^3)\right],
 \label{eq:wkb_residue_exact}\\
 \delta_A^{(2)}&=e^{-BM_A}
 \left[\partial_z\Delta_A^{(2)}-2B\Delta_A^{(2)}\right]_{z=M_A}.
 \label{eq:wkb_delta_residue}
\end{align}
Thus $|\delta_A^{(2)}|<1$ is a sufficient positivity condition. For the evolving-history benchmark in Appendix~\ref{app:closure}, $\max|\delta_A^{(2)}|=3.28\times10^{-3}$ over $H\le p\le50H$, so $Z_A>0.9967e^{-BM_A}>0$. In the switching layers, where an instantaneous-pole picture is least reliable, the exact norm relation \eqref{eq:symplectic_congruence} gives the relevant result rather than the WKB approximation.

The activated phase is described by the invariant potential
\begin{equation}
 V_Q(Q,C)
 =\frac12\mu^2(C)I_2-\frac13\gamma(C)I_3+\frac14\lambda(C)I_2^2,
 \qquad
 I_2=Q_{\mu\nu}Q^{\mu\nu},
 \quad
 I_3=Q_\mu{}^\nu Q_\nu{}^\rho Q_\rho{}^\mu.
 \label{eq:main_phase_potential}
\end{equation}
For $\lambda>0$ and discriminant $\gamma^2-24\lambda\mu^2\ge0$, the nonzero uniaxial amplitude is
\begin{equation}
 D(C)=\frac{\gamma(C)+\sqrt{\gamma^2(C)-24\lambda(C)\mu^2(C)}}{4\lambda(C)}.
 \label{eq:main_D_amplitude}
\end{equation}
The plus branch is selected by the positive Hessian condition. The cubic invariant is physically decisive: it distinguishes uniaxial eigenvalue patterns and fixes the sign of the ordered phase, rather than leaving a continuous sphere of degenerate tensor amplitudes.

The stationary-point classification and branch selection show explicitly why the preferred configuration is a dynamically selected uniaxial phase rather than an imposed ansatz:

\paragraph*{Scalar invariants of the tracefree order parameter}

The field $Q_{\mu\nu}$ is self-adjoint with respect to the positive spatial metric $h_{\mu\nu}$. At each spacetime point it is therefore diagonalizable by an orthonormal spatial triad. Let its eigenvalues be
\begin{equation}
  q_1,\quad q_2,\quad q_3,
  \qquad
  q_1+q_2+q_3=0 .
  \label{mph:eq:Q_eigenvalues_section4}
\end{equation}
The trace condition is the last equality. Rotationally invariant scalar functions of $Q_{\mu\nu}$ are generated by traces of powers of the mixed tensor
\begin{equation}
  Q_\mu{}^\nu
  =
  h^{\nu\rho}Q_{\mu\rho}.
\label{Xeqn71-88}
\end{equation}
Because $Q_\mu{}^\mu=0$, the first nonzero invariant is quadratic:
\begin{equation}
  I_2
  \equiv
  Q_{\mu\nu}Q^{\mu\nu}
  =
  {\rm Tr}(Q^2)
  =
  q_1^2+q_2^2+q_3^2 .
  \label{mph:eq:I2_definition_section4}
\end{equation}
Since $h_{\mu\nu}$ is positive definite on the spatial hypersurfaces, $I_2\geq0$, with equality precisely at $Q_{\mu\nu}=0$.

The independent cubic invariant is
\begin{equation}
  I_3
  \equiv
  Q_\mu{}^\nu Q_\nu{}^\rho Q_\rho{}^\mu
  =
  {\rm Tr}(Q^3)
  =
  q_1^3+q_2^3+q_3^3 .
  \label{mph:eq:I3_definition_section4}
\end{equation}
Using $q_1+q_2+q_3=0$, the algebraic identity
\begin{equation}
  q_1^3+q_2^3+q_3^3
  =
  3q_1q_2q_3
  \label{mph:eq:I3_product_identity_section4}
\end{equation}
gives
\begin{equation}
  I_3=3q_1q_2q_3 .
\label{Xeqn75-92}
\end{equation}

The Cayley--Hamilton theorem for the tracefree $3\times3$ matrix $Q_\mu{}^\nu$ gives
\begin{equation}
  Q^3
  -
  \frac12I_2 Q
  -
  \frac13I_3 h
  =
  0 ,
  \label{mph:eq:Cayley_Hamilton_section4}
\end{equation}
where $h$ denotes the identity map on the spatial tangent space. Multiplying Eq.~\eqref{mph:eq:Cayley_Hamilton_section4} by $Q$ and taking the trace gives
\begin{equation}
  {\rm Tr}(Q^4)
  =
  \frac12 I_2^2 .
  \label{mph:eq:quartic_identity_section4}
\end{equation}
Thus there is only one independent quartic invariant. A local analytic scalar potential for the tracefree order parameter through quartic order is therefore a function of $I_2$, $I_3$, and $I_2^2$.

\paragraph*{Curvature-history phase potential}

The order-parameter part of the scalar potential in Eq.~\eqref{eq:main_phase_potential} is specified by the invariant polynomial
\begin{equation}
  V_Q(Q,C)
  =
  V_0(C)
  +
  \frac12\mu^2(C)I_2
  -
  \frac13\gamma(C)I_3
  +
  \frac14\lambda(C)I_2^2 .
  \label{mph:eq:VQ_definition_section4}
\end{equation}
The functions $V_0(C)$, $\mu^2(C)$, $\gamma(C)$, and $\lambda(C)$ are scalar functions of the curvature-history field. The quartic coefficient is positive,
\begin{equation}
  \lambda(C)>0,
  \label{mph:eq:lambda_positive_section4}
\end{equation}
so that the potential is bounded from below at large $I_2$. The cubic coefficient is chosen positive,
\begin{equation}
  \gamma(C)>0,
  \label{mph:eq:gamma_positive_section4}
\end{equation}
which fixes the prolate orientation of the uniaxial phase. The sign of the axis is still redundant because the final tensor is invariant under $n^\mu\mapsto -n^\mu$.

The derivative of $V_Q$ with respect to $Q_{\mu\nu}$ is computed inside the spatial symmetric tracefree subspace. The quadratic invariant varies as
\begin{equation}
  \delta I_2
  =
  2Q^{\mu\nu}\delta Q_{\mu\nu}.
  \label{mph:eq:delta_I2_section4}
\end{equation}
The cubic invariant varies as
\begin{align}
  \delta I_3
  &=
  3Q_\mu{}^\rho Q_\rho{}^\nu\delta Q_\nu{}^\mu .
\end{align}
Since $\delta Q_{\mu\nu}$ is tracefree, only the tracefree part of $Q_\mu{}^\rho Q_{\rho\nu}$ contributes. Therefore
\begin{equation}
  \delta I_3
  =
  3
  \left(
    Q_\mu{}^\rho Q_{\rho\nu}
    -
    \frac13I_2h_{\mu\nu}
  \right)
  \delta Q^{\mu\nu}.
  \label{mph:eq:delta_I3_section4}
\end{equation}
The variation of Eq.~\eqref{mph:eq:VQ_definition_section4} is then
\begin{equation}
  \delta V_Q
  =
  \mathcal{F}_{\mu\nu}\delta Q^{\mu\nu},
  \label{mph:eq:delta_VQ_section4}
\end{equation}
where
\begin{equation}
  \mathcal{F}_{\mu\nu}
  =
  \left[
    \mu^2(C)+\lambda(C)I_2
  \right]Q_{\mu\nu}
  -
  \gamma(C)
  \left(
    Q_\mu{}^\rho Q_{\rho\nu}
    -
    \frac13I_2h_{\mu\nu}
  \right).
  \label{mph:eq:F_tensor_section4}
\end{equation}
The phase extrema are therefore the solutions of
\begin{equation}
  \mathcal{F}_{\mu\nu}=0.
  \label{mph:eq:Q_stationarity_tensor_section4}
\end{equation}

\paragraph*{Algebraic classification of stationary phases}

At a point, choose an orthonormal spatial triad that diagonalizes $Q_\mu{}^\nu$. In this frame Eq.~\eqref{mph:eq:Q_stationarity_tensor_section4} becomes three scalar equations:
\begin{equation}
  \left[
    \mu^2+\lambda I_2
  \right]q_a
  -
  \gamma
  \left(
    q_a^2-\frac13I_2
  \right)
  =
  0,
  \qquad
  a=1,2,3 .
  \label{mph:eq:eigenvalue_stationarity_section4}
\end{equation}
Subtracting the equation for $q_b$ from the equation for $q_a$ gives
\begin{equation}
  (q_a-q_b)
  \left[
    \mu^2+\lambda I_2-\gamma(q_a+q_b)
  \right]
  =
  0 .
  \label{mph:eq:eigenvalue_difference_section4}
\end{equation}

A nonzero stationary solution cannot have three distinct eigenvalues. Indeed, for three distinct eigenvalues Eq.~\eqref{mph:eq:eigenvalue_difference_section4} gives
\begin{equation}
  q_1+q_2
  =
  q_1+q_3
  =
  q_2+q_3
  =
  \frac{\mu^2+\lambda I_2}{\gamma}.
\label{Xeqn88-106}
\end{equation}
Subtracting the first two equalities gives $q_2=q_3$, contradicting distinctness. Therefore every nonzero stationary solution has two equal eigenvalues.

Let the repeated eigenvalue be $-D/3$ and the remaining eigenvalue be $2D/3$. The trace condition is then automatic:
\begin{equation}
  \frac{2D}{3}
  -
  \frac{D}{3}
  -
  \frac{D}{3}
  =
  0 .
\label{Xeqn89-107}
\end{equation}
There exists a spatial unit vector $n^\mu$ along the nondegenerate eigenvector,
\begin{equation}
  n^\mu u_\mu=0,
  \qquad
  n^\mu n_\mu=1 .
  \label{mph:eq:n_unit_section4}
\end{equation}
The corresponding tensor is
\begin{equation}
  Q_{\mu\nu}
  =
  D
  \left(
    n_\mu n_\nu
    -
    \frac13h_{\mu\nu}
  \right).
  \label{mph:eq:uniaxial_Q_section4}
\end{equation}
Eq.~\eqref{mph:eq:uniaxial_Q_section4} has eigenvalue $2D/3$ along $n^\mu$ and eigenvalue $-D/3$ on the two-dimensional plane orthogonal to both $u^\mu$ and $n^\mu$. It is symmetric, spatial, and tracefree:
\begin{equation}
  Q_{\mu\nu}u^\nu=0,
  \qquad
  h^{\mu\nu}Q_{\mu\nu}
  =
  D
  \left(
    h^{\mu\nu}n_\mu n_\nu
    -
    \frac13h^{\mu\nu}h_{\mu\nu}
  \right)
  =
  D(1-1)
  =
  0 .
\label{Xeqn92-110}
\end{equation}

The residual symmetry group of Eq.~\eqref{mph:eq:uniaxial_Q_section4} is the group of rotations preserving $n^\mu$. Rotations in the two-plane orthogonal to $n^\mu$ leave both $n_\mu n_\nu$ and $h_{\mu\nu}$ invariant. Hence the tensor-memory phase breaks the spatial rotation group according to
\begin{equation}
  SO(3)\longrightarrow SO(2).
  \label{mph:eq:symmetry_breaking_section4}
\end{equation}

\paragraph*{Activation amplitude}

For the uniaxial tensor \eqref{mph:eq:uniaxial_Q_section4}, the invariants are
\begin{equation}
  I_2
  =
  \left(\frac{2D}{3}\right)^2
  +
  2\left(-\frac{D}{3}\right)^2
  =
  \frac23D^2,
  \label{mph:eq:I2_uniaxial_section4}
\end{equation}
and
\begin{equation}
  I_3
  =
  \left(\frac{2D}{3}\right)^3
  +
  2\left(-\frac{D}{3}\right)^3
  =
  \frac29D^3 .
  \label{mph:eq:I3_uniaxial_section4}
\end{equation}
Substitution into Eq.~\eqref{mph:eq:VQ_definition_section4} gives the one-variable phase potential
\begin{equation}
  V_Q(D,C)
  =
  V_0(C)
  +
  \frac13\mu^2(C)D^2
  -
  \frac{2}{27}\gamma(C)D^3
  +
  \frac19\lambda(C)D^4 .
  \label{mph:eq:VQ_D_section4}
\end{equation}
Its derivative with respect to $D$ is
\begin{align}
  \frac{\partial V_Q}{\partial D}
  &=
  \frac23\mu^2D
  -
  \frac29\gamma D^2
  +
  \frac49\lambda D^3
  \notag\\
  &=
  \frac{2D}{9}
  \left(
    3\mu^2-\gamma D+2\lambda D^2
  \right).
  \label{mph:eq:dVdD_section4}
\end{align}
Thus the nonzero stationary amplitudes obey
\begin{equation}
  2\lambda(C)D^2
  -
  \gamma(C)D
  +
  3\mu^2(C)
  =
  0 .
  \label{mph:eq:D_quadratic_section4}
\end{equation}
The two algebraic roots are
\begin{equation}
  D_\pm(C)
  =
  \frac{
    \gamma(C)
    \pm
    \sqrt{
      \gamma^2(C)
      -
      24\lambda(C)\mu^2(C)
    }
  }{
    4\lambda(C)
  } .
  \label{mph:eq:D_roots_section4}
\end{equation}
The activated branch is the stable nonzero root
\begin{equation}
  D(C)
  =
  D_+(C)
  =
  \frac{
    \gamma(C)
    +
    \sqrt{
      \gamma^2(C)
      -
      24\lambda(C)\mu^2(C)
    }
  }{
    4\lambda(C)
  } .
  \label{mph:eq:D_activated_section4}
\end{equation}
The stability of this branch follows from the second derivative. Differentiating Eq.~\eqref{mph:eq:dVdD_section4} gives
\begin{equation}
  \frac{\partial^2V_Q}{\partial D^2}
  =
  \frac23\mu^2
  -
  \frac49\gamma D
  +
  \frac43\lambda D^2 .
\label{Xeqn100-119}
\end{equation}
Using the stationary relation $3\mu^2=\gamma D-2\lambda D^2$, this becomes
\begin{equation}
  \left.
  \frac{\partial^2V_Q}{\partial D^2}
  \right|_{D=D_\pm}
  =
  \frac{2D_\pm}{9}
  \left(
    4\lambda D_\pm-\gamma
  \right).
  \label{mph:eq:second_derivative_D_section4}
\end{equation}
For $D_+$,
\begin{equation}
  4\lambda D_+-\gamma
  =
  \sqrt{
    \gamma^2-24\lambda\mu^2
  },
\label{Xeqn102-121}
\end{equation}
so Eq.~\eqref{mph:eq:second_derivative_D_section4} is positive on the activated branch.

The equality of the isotropic value $D=0$ and the nonzero extremum occurs when
\begin{equation}
  V_Q(D,C)=V_Q(0,C).
  \label{mph:eq:degenerate_condition_section4}
\end{equation}
Using Eq.~\eqref{mph:eq:VQ_D_section4} and the stationary relation \eqref{mph:eq:D_quadratic_section4}, the potential difference at a nonzero extremum is
\begin{equation}
  V_Q(D,C)-V_0(C)
  =
  \frac{D^3}{27}
  \left[
    \gamma(C)-3\lambda(C)D
  \right].
  \label{mph:eq:V_difference_extremum_section4}
\end{equation}
Degeneracy therefore occurs at
\begin{equation}
  D_{\rm c}(C)
  =
  \frac{\gamma(C)}{3\lambda(C)} .
  \label{mph:eq:Dcritical_section4}
\end{equation}
Substitution of Eq.~\eqref{mph:eq:Dcritical_section4} into Eq.~\eqref{mph:eq:D_quadratic_section4} gives the curvature-history activation equation
\begin{equation}
  \mu^2(C)
  =
  \frac{\gamma^2(C)}{27\lambda(C)} .
  \label{mph:eq:activation_equation_section4}
\end{equation}
The activated domain is the region of curvature-history field space in which the nonzero uniaxial extremum has lower potential energy than the isotropic point:
\begin{equation}
  \mathfrak{A}
  =
  \left\{
    C\; \Big|\;
    \mu^2(C)
    <
    \frac{\gamma^2(C)}{27\lambda(C)}
  \right\}.
  \label{mph:eq:activated_domain_section4}
\end{equation}
For $C\in\mathfrak{A}$, the order parameter is
\begin{equation}
  Q_{\mu\nu}^{\rm act}
  =
  D(C)
  \left(
    n_\mu n_\nu
    -
    \frac13h_{\mu\nu}
  \right),
  \qquad
  D(C)=D_+(C).
  \label{mph:eq:activated_Q_section4}
\end{equation}

\paragraph*{Biaxial exclusion and uniqueness of the uniaxial minimum}

The uniaxial form \eqref{mph:eq:activated_Q_section4} is not a gauge choice. It follows from the invariant algebra of a tracefree spatial tensor. For real eigenvalues satisfying $q_1+q_2+q_3=0$, the characteristic polynomial is
\begin{equation}
  p(q)
  =
  q^3-\frac12I_2q-\frac13I_3 .
  \label{mph:eq:characteristic_polynomial_section4}
\end{equation}
The discriminant of this cubic is nonnegative because all three eigenvalues are real. This gives
\begin{equation}
  \frac12 I_2^3-3I_3^2\geq0,
\label{Xeqn110-129}
\end{equation}
or equivalently
\begin{equation}
  6I_3^2\leq I_2^3 .
  \label{mph:eq:biaxial_inequality_section4}
\end{equation}
Equality holds precisely when the cubic has a repeated root, which is exactly the uniaxial case.

At fixed $I_2$, the quadratic and quartic terms in Eq.~\eqref{mph:eq:VQ_definition_section4} are fixed. The angular dependence of the potential is entirely contained in the term
\begin{equation}
  -\frac13\gamma I_3 .
\label{Xeqn112-131}
\end{equation}
Since $\gamma>0$, minimizing the potential at fixed $I_2$ requires maximizing $I_3$. By Eq.~\eqref{mph:eq:biaxial_inequality_section4}, the maximum value is
\begin{equation}
  I_3
  =
  \frac{1}{\sqrt{6}}I_2^{3/2},
  \label{mph:eq:I3_max_section4}
\end{equation}
and this value is attained precisely by the uniaxial spectrum
\begin{equation}
  \left(
    \frac{2D}{3},
    -\frac{D}{3},
    -\frac{D}{3}
  \right).
\label{Xeqn114-133}
\end{equation}
Therefore the activated minimum of the phase potential is uniaxial and is represented uniquely, up to spatial rotations and the sign redundancy $n^\mu\mapsto -n^\mu$, by Eq.~\eqref{mph:eq:activated_Q_section4}.

Local scalar stability is controlled by the reduced curvature action derived later. The spatial gradient matrix is
\begin{equation}
 \Gamma^{ij}=\delta^{ij}-bP^{ij},
 \qquad
 P^{ij}=\hat n^i\hat n^j-\frac13\delta^{ij}.
 \label{eq:main_gradient_matrix}
\end{equation}
Its eigenvalue parallel to $\hat{\mathbf n}$ is $1-2b/3$, while each transverse eigenvalue is $1+b/3$. Positivity therefore requires
\begin{equation}
 -3<b<\frac32.
 \label{eq:main_stability_interval}
\end{equation}
This interval is a direct dynamical constraint on the memory coupling, not an observational prior. At either endpoint, one gradient eigenvalue vanishes and the effective scalar description becomes strongly coupled; outside it, the corresponding mode has a gradient instability.

The closure equations and Einstein constraints also preserve the standard number of gravitational tensor modes. Lapse and shift remain Lagrange multipliers, and the auxiliary tensor components are removed algebraically. The new physical content is therefore limited to the five-component memory order parameter and its effect on scalar perturbations. Appendix~\ref{app:closure} gives the full constraint count, Hamiltonian positivity test, and Hessian eigenvalues.

\section{Homogeneous inflationary dynamics}\label{Xsec8-5}
\label{sec:main_background}

For the spatially flat FLRW geometry,
\begin{equation}
 ds^2=-dt^2+a^2(t)\delta_{ij}dx^idx^j,
 \qquad
 H=\frac{\dot a}{a},
 \qquad
 \epsilon=-\frac{\dot H}{H^2}.
 \label{eq:main_FLRW}
\end{equation}
The inflaton-defined normal is $u^\mu=(1,0,0,0)$ on the homogeneous branch.  The Gauss--Bonnet history source is
\begin{equation}
 \mathcal G_{\rm GB}=24H^2(H^2+\dot H)=24H^4(1-\epsilon).
 \label{eq:main_GB_FLRW}
\end{equation}
For comparison, the Kretschmann scalar is $24H^4(1-\epsilon+\epsilon^2/2)$, so the completion changes the slow-roll source only at $O(\epsilon^2)$ while eliminating the nonlinear curvature-squared Ostrogradsky sector \cite{bib0039}.

For homogeneous fields, Eq.~\eqref{eq:main_C_eom} becomes
\begin{equation}
 \ddot{\bar C}+3H\dot{\bar C}+m_c^2\bar C
 =24\ell_*H^4(1-\epsilon).
 \label{eq:main_C_background}
\end{equation}
The three terms on the left have direct physical meanings: inertial response, Hubble damping, and relaxation toward zero history in the absence of curvature. In exact de Sitter space, the stationary solution is
\begin{equation}
 C_0=\frac{24\ell_*H_*^4}{m_c^2}.
 \label{eq:main_C0}
\end{equation}
Thus the parameter entering the phase potential is fixed by the inflationary curvature scale and the memory range; it is not an independent constant introduced solely to fit the anisotropy.

The conjugate response obeys
\begin{equation}
 \ddot{\bar\chi}+3H\dot{\bar\chi}+m_c^2\bar\chi=\bar\Sigma_\Delta.
 \label{eq:main_chi_background}
\end{equation}
Together with the inflaton equation and the $00$ Einstein equation, the homogeneous system may be written as
\begin{align}
 3M_{\rm Pl}^2H^2&=\frac12\dot{\bar\phi}^{\,2}+U(\bar\phi)+\rho_{C\Delta},\\
 -2M_{\rm Pl}^2\dot H&=\dot{\bar\phi}^{\,2}+\rho_{C\Delta}+p_{C\Delta},\\
 \ddot{\bar\phi}+3H\dot{\bar\phi}+U_{,\phi}&=\bar{\mathcal J}_\phi^{(\Delta)},
 \label{eq:main_background_Einstein}
\end{align}
where $\rho_{C\Delta}$ and $p_{C\Delta}$ are projections of the coupled memory--history stress tensor. Their invariant definitions and the background reduction of the coupled history--memory source are summarized in Appendix~\ref{app:background_activation}.

The background has two regimes. Before activation, $Q_{\mu\nu}=0$ and the memory sector remains subdominant, so standard slow roll gives the leading background. When $C(t)$ enters the region where the nonzero minimum of $V_Q$ is preferred, the tensor order parameter develops a finite amplitude. Because $Q_{\mu\nu}$ is tracefree, the homogeneous energy density can stay nearly isotropic even though perturbations respond to its principal axis through gradient terms. This allows weak statistical anisotropy without a large Bianchi-type deformation of the background.

Activation is dynamical. The amplitude obeys the projected memory equation and follows the instantaneous minimum only when its response time is short compared with the evolution time of $C$. Thus, \eqref{eq:main_D_amplitude} gives an equilibrium branch, not a substitute for the time-dependent field equation. In the finite-window approximation, the active interval represents the period during which the background amplitude is close enough to the nonzero branch to affect scalar modes.

\subsection{Representative dynamical activation profile}\label{Xsec9-5.1}
\label{subsec:explicit_activation}

To demonstrate that the finite window can arise dynamically, take constant $\gamma(C)=\gamma_0$ and $\lambda(C)=\lambda_0>0$ and choose a smooth finite band in the quadratic coefficient,
\begin{align}
 \mu^2(C)&=\mu_{\rm out}^2-\Delta\mu^2\,W_C(C),
 \label{eq:mu_window}\\
 W_C(C)&=\frac12\left[
 \tanh\frac{C-C_-}{\sigma_C}
 -\tanh\frac{C-C_+}{\sigma_C}\right],
 \qquad C_-<C_+.
 \label{eq:C_window}
\end{align}
For $C_-+O(\sigma_C)<C<C_+-O(\sigma_C)$, $W_C\simeq1$; outside the band $W_C\simeq0$.  A concrete regular benchmark is
\begin{equation}
 \gamma_0=6H_*,\quad \lambda_0=1,\quad
 \mu_{\rm out}^2=9H_*^2,\quad
 \mu_{\rm in}^2\equiv\mu_{\rm out}^2-\Delta\mu^2=-4H_*^2,
 \quad m_c=4H_*.
 \label{eq:activation_benchmark_parameters}
\end{equation}
Outside the band, $\gamma_0^2-24\lambda_0\mu_{\rm out}^2<0$, so no nonzero stationary branch exists.  Inside, $D=0$ is tachyonic and the stable branch is
\begin{equation}
 D_+=\frac{\gamma_0+\sqrt{\gamma_0^2-24\lambda_0\mu_{\rm in}^2}}{4\lambda_0}
 =\frac{6+\sqrt{132}}{4}H_*\simeq4.372H_*.
 \label{eq:D_plateau_value}
\end{equation}

For this background demonstration we also take $B(C)=B_0$ and $G(C)=G_0$ across the narrow phase band; all activation is then carried by $\mu^2(C)$.  Since $I_2=2D^2/3$, the retarded response equation is
\begin{equation}
 \ddot\chi+3H\dot\chi+m_c^2\chi
 =-\frac13\mu^2_{,C}(C)D^2,
 \qquad
 \chi|_{\Sigma_*}=\dot\chi|_{\Sigma_*}=0.
 \label{eq:chi_activation_benchmark}
\end{equation}

For the uniaxial ansatz the canonically normalized amplitude equation obtained from the dressed local action is
\begin{equation}
 \ddot D+3H\dot D+\mu^2(C)D-\frac{\gamma_0}{3}D^2+\frac{2\lambda_0}{3}D^3=0.
 \label{eq:D_dynamic}
\end{equation}
Linearizing about $D_+(C)$ gives
\begin{equation}
 \delta\ddot D+3H\delta\dot D+m_D^2(C)\delta D
 =-\left(\ddot D_++3H\dot D_+\right),
 \qquad
 m_D^2(C)=\frac{D_+}{3}\sqrt{\gamma_0^2-24\lambda_0\mu^2(C)}.
 \label{eq:D_tracking}
\end{equation}
For the benchmark, $m_D^2\simeq16.75H_*^2$, so deviations from the plateau are underdamped with an envelope $|\delta D|\propto a^{-3/2}$ once the branch has been reached.

For approximately constant slow-roll parameter, $H(N)=H_*e^{-\epsilon(N-N_*)}$ and the retarded history field tracks
\begin{equation}
 C_{\rm eq}(N)=\frac{24\ell_*}{m_c^2}H^4(N)(1-\epsilon),
 \qquad
 \frac{C-C_{\rm eq}}{C_{\rm eq}}
 =O\!\left(\epsilon\frac{H^2}{m_c^2}\right).
 \label{eq:C_adiabatic_solution}
\end{equation}
Hence $C(N)$ is monotone and the two band crossings are predicted rather than inserted:
\begin{equation}
 N_i-N_* =\frac{1}{4\epsilon}\ln\frac{C_*}{C_+},
 \qquad
 N_f-N_* =\frac{1}{4\epsilon}\ln\frac{C_*}{C_-},
 \qquad
 \Delta N=\frac{1}{4\epsilon}\ln\frac{C_+}{C_-}.
 \label{eq:predicted_window_efolds}
\end{equation}
For $\epsilon=10^{-2}$, $C_+=1.17C_a$ and $C_-=0.75C_a$, and Eq.~\eqref{eq:predicted_window_efolds} gives $\Delta N\simeq11.1$. With $C(0)=1.18C_a$, the numerical solution crosses the band first at $N\simeq0.3$ and again at $N\simeq11.5$. Choosing $\sigma_C=0.02C_a$ makes each switching region much shorter than the plateau. The conformal-time endpoints follow from $\tau\simeq-[(1-\epsilon)aH]^{-1}$, so $\tau_i$ and $\tau_f$ are fixed by the microscopic band parameters.

\begin{figure}
\centering
\includegraphics[width=0.80\linewidth]{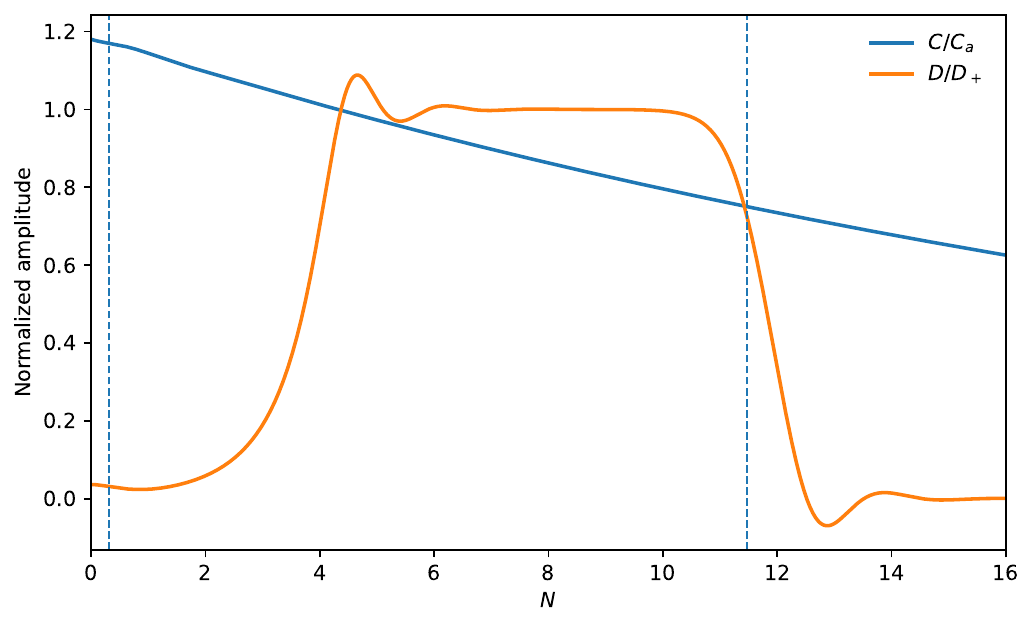}
\caption{Representative dynamical activation.  The retarded history response $C(N)$ crosses the smooth band \eqref{eq:C_window}; the order-parameter amplitude $D(N)$ grows from a fluctuation seed, relaxes toward $D_+$, remains approximately constant over the interior of the band, and decays after exit.  The plotted quantities are normalized to their plateau values.}
\label{fig:activation_dynamics}
\end{figure}

For this benchmark, and at zeroth order in the weak Ricci-coupling parameter $b$, the homogeneous response and uniaxial phase energies are
\begin{align}
 \rho_C&=\dot\chi\dot C+m_c^2\chi C-24\ell_*H^3\dot\chi,
 \label{eq:rho_C_benchmark}\\
 \rho_D&=\frac13\dot D^2+V_Q(D,C).
 \label{eq:rho_D_benchmark}
\end{align}
The numerical solution in units $H_*=C_a=1$ gives
\begin{equation}
 N_i=0.3056,\qquad N_f=11.4773,\qquad
 \max_{6\le N\le10}\left|\frac{D}{D_+}-1\right|<9.2\times10^{-3},
 \label{eq:numerical_activation_diagnostics}
\end{equation}
and $\max|\rho_C+\rho_D|<2.76\times10^3H_*^4$.  Hence, for the conservative inflationary scale $H_*/M_{\rm Pl}\le10^{-5}$,
\begin{equation}
 \Omega_{\rm mem}
 \equiv\frac{|\rho_{C\Delta}|}{3M_{\rm Pl}^2H^2}
 <1.2\times10^{-7}+O(|b|\epsilon)\ll1.
 \label{eq:memory_subdominant}
\end{equation}
The use of the zeroth-order slow-roll $H(N)$ in the response integration is therefore self-consistent rather than circular.  Writing $H^2=H_{\rm sr}^2(1+\delta_H)$ in the exact Friedmann equation gives
\begin{equation}
 |\delta_H|\le \Omega_{\rm mem}+O(\Omega_{\rm mem}^2),
 \qquad
 \left|\frac{\delta H}{H_{\rm sr}}\right|
 \le\frac12\Omega_{\rm mem}+O(\Omega_{\rm mem}^2)
 <6\times10^{-8}.
 \label{eq:H_backreaction_bound}
\end{equation}
Standard continuous-dependence results for ordinary differential equations imply that using the self-consistent Hubble rate changes $C(N)$, $\chi(N)$, and $D(N)$ by only $O(10^{-7})$ over the interval shown in Fig.~\ref{fig:activation_dynamics}. This is much smaller than the percent-level plateau diagnostic in \eqref{eq:numerical_activation_diagnostics}. The finite window is therefore a controlled approximation to the coupled solution, not an externally imposed top-hat.
The same parameter choice therefore gives a finite dynamical activation, an approximately constant four-e-fold plateau, and negligible backreaction on slow roll. Fig.~\ref{fig:activation_dynamics} shows a numerical integration of Eqs.~\eqref{eq:main_C_background}, \eqref{eq:chi_activation_benchmark}, and \eqref{eq:D_dynamic}. The classical instability starts from the standard fluctuation-size seed $D(0)=H_*/(2\pi)$. The plateau is insensitive to the precise nonzero seed because the tachyonic branch is an attractor.

\section{Tensor-memory activation and primordial dynamics}\label{Xsec10-6}
\label{sec:main_perturbations}

The spatial tracefree projector is
\begin{equation}
 \Pi_{\mu\nu}{}^{\alpha\beta}
 =h_{(\mu}{}^\alpha h_{\nu)}{}^\beta-\frac13h_{\mu\nu}h^{\alpha\beta}.
 \label{eq:main_STF_projector}
\end{equation}
At a nonzero stationary point, the invariant-potential equations select a stable uniaxial phase. It can be written as Eq.~\eqref{eq:main_uniaxial_intro}, with eigenvalues $(2D/3,-D/3,-D/3)$. The remaining symmetry is $SO(2)$ about $n^\mu$. The phase selects an axis, not an oriented vector: $n^\mu$ and $-n^\mu$ describe the same tensor. This is why the leading scalar signal has even parity.

To derive the scalar observable, we use the ADM decomposition \cite{bib0008,bib0009,bib0010},
\begin{equation}
 ds^2=-N^2dt^2+h_{ij}(dx^i+N^idt)(dx^j+N^jdt)
 \label{eq:main_ADM_metric}
\end{equation}
and comoving gauge,
\begin{equation}
 \delta\phi=0,
 \qquad
 h_{ij}=a^2e^{2\zeta}(e^\gamma)_{ij},
 \qquad
 \partial_i\gamma_{ij}=\gamma_{ii}=0.
 \label{eq:main_comoving_gauge}
\end{equation}
For the scalar sector, write $N=1+\alpha$ and $N_i=\partial_i\beta+N_i^T$. Functional differentiation of the full memory--history action gives the scalar source variables $\delta\mathcal E_{\rm mem}$ and $\mathcal P_{\rm mem}$. Before imposing the activated-branch simplifications, the constraint derivation is:

\paragraph*{Scalar perturbative ADM variables}

The scalar perturbative ADM variables are introduced by
\begin{equation}
  N
  =
  1+\alpha,
  \qquad
  N_i
  =
  \partial_i\beta,
  \qquad
  \gamma_{ij}
  =
  a^2(t)e^{2\zeta}\delta_{ij}.
  \label{madm:eq:scalar_ADM_variables_section6}
\end{equation}
The inflaton is represented in comoving slicing by
\begin{equation}
  \phi(t,\mathbf{x})=\bar\phi(t).
  \label{madm:eq:comoving_phi_section6}
\end{equation}
All expressions in this subsection are kept to first order in perturbations.

The inverse spatial metric is
\begin{equation}
  \gamma^{ij}
  =
  a^{-2}(1-2\zeta)\delta^{ij}.
  \label{madm:eq:gamma_inverse_linear_section6}
\end{equation}
The three-curvature of $\gamma_{ij}=a^2e^{2\zeta}\delta_{ij}$ is
\begin{equation}
  {}^{(3)}R
  =
  -4a^{-2}\partial^2\zeta,
  \label{madm:eq:R3_linear_section6}
\end{equation}
where $\partial^2=\delta^{ij}\partial_i\partial_j$. This follows from the conformal transformation of the Ricci scalar in three dimensions,
\begin{equation}
  {}^{(3)}R[e^{2\zeta}\delta]
  =
  e^{-2\zeta}
  \left[
    -4\partial^2\zeta
    -
    2(\partial_i\zeta)(\partial^i\zeta)
  \right],
\label{Xeqn139-165}
\end{equation}
and retention of the linear term.

Using Eq.~\eqref{eq:K_ADM_section6}, the mixed extrinsic curvature is
\begin{equation}
  K^i{}_j
  =
  H\delta^i{}_j
  +
  \left(
    \dot\zeta-H\alpha
  \right)\delta^i{}_j
  -
  a^{-2}\partial^i\partial_j\beta.
  \label{madm:eq:Kij_linear_section6}
\end{equation}
Its trace is
\begin{equation}
  K
  =
  3H
  +
  3(\dot\zeta-H\alpha)
  -
  a^{-2}\partial^2\beta .
  \label{madm:eq:K_linear_section6}
\end{equation}
Thus,
\begin{align}
  K^2-K_{ij}K^{ij}
  &=
  6H^2
  +
  4H
  \left[
    3(\dot\zeta-H\alpha)
    -
    a^{-2}\partial^2\beta
  \right].
  \label{madm:eq:K_combination_linear_section6}
\end{align}

The normal derivative of the comoving inflaton is
\begin{equation}
  \Pi_\phi
  =
  \frac{\dot{\bar\phi}}{1+\alpha}
  =
  \dot{\bar\phi}
  -
  \alpha\dot{\bar\phi}.
  \label{madm:eq:Pi_phi_linear_section6}
\end{equation}
Since $D_i\phi=0$ in comoving slicing,
\begin{equation}
  \delta\mathcal{P}_i^{(\phi)}=0.
  \label{madm:eq:delta_P_phi_zero_section6}
\end{equation}
The inflaton energy perturbation is
\begin{equation}
  \delta\mathcal{E}_\phi
  =
  -\dot{\bar\phi}^{\,2}\alpha .
  \label{madm:eq:delta_E_phi_section6}
\end{equation}

The scalar part of the memory--history ADM momentum density is written as the gradient of a scalar potential:
\begin{equation}
  \delta\mathcal{P}^{({\rm mem})}_i
  =
  \partial_i\mathcal{P}_{\rm mem}.
  \label{madm:eq:Pmem_scalar_potential_section6}
\end{equation}
The corresponding energy perturbation is denoted by
\begin{equation}
  \delta\mathcal{E}_{\rm mem}
  =
  \mathcal{E}_{\rm mem}
  -
  \bar{\mathcal{E}}_{\rm mem}.
  \label{madm:eq:delta_E_mem_section6}
\end{equation}

\paragraph*{Linear Hamiltonian constraint}

Subtracting the background Hamiltonian constraint \eqref{eq:background_Hamiltonian_section6} from the exact constraint \eqref{eq:exact_H_constraint_total_section6} and using Eqs.~\eqref{madm:eq:R3_linear_section6} -- \eqref{madm:eq:K_combination_linear_section6} gives
\begin{align}
  -4a^{-2}\partial^2\zeta
  +
  12H\dot\zeta
  -
  12H^2\alpha
  -
  4Ha^{-2}\partial^2\beta
  =
  \frac{2}{M_{\rm Pl}^2}
  \left(
    -\dot{\bar\phi}^{\,2}\alpha
    +
    \delta\mathcal{E}_{\rm mem}
  \right).
  \label{madm:eq:linear_H_constraint_raw_section6}
\end{align}
Dividing by four gives
\begin{equation}
  -a^{-2}\partial^2\zeta
  +
  3H\dot\zeta
  -
  3H^2\alpha
  -
  Ha^{-2}\partial^2\beta
  =
  \frac{1}{2M_{\rm Pl}^2}
  \left(
    -\dot{\bar\phi}^{\,2}\alpha
    +
    \delta\mathcal{E}_{\rm mem}
  \right).
  \label{madm:eq:linear_H_constraint_section6}
\end{equation}

Solving Eq.~\eqref{madm:eq:linear_H_constraint_section6} for the scalar shift gives
\begin{equation}
  a^{-2}\partial^2\beta
  =
  -\frac{1}{H}a^{-2}\partial^2\zeta
  +
  3\dot\zeta
  -
  3H\alpha
  +
  \frac{\dot{\bar\phi}^{\,2}}{2M_{\rm Pl}^2H}\alpha
  -
  \frac{1}{2M_{\rm Pl}^2H}\delta\mathcal{E}_{\rm mem}.
  \label{madm:eq:beta_constraint_unsolved_alpha_section6}
\end{equation}
Using the exact background relation
\begin{equation}
  \epsilon
  =
  -\frac{\dot H}{H^2}
  =
  \frac{
    \dot{\bar\phi}^{\,2}
    +
    \bar{\mathcal{E}}_{\rm mem}
    +
    \bar p_{\rm mem}
  }{
    2M_{\rm Pl}^2H^2
  },
  \label{madm:eq:epsilon_background_section6}
\end{equation}
the coefficient of $\alpha$ may also be written as
\begin{equation}
  -3H
  +
  \frac{\dot{\bar\phi}^{\,2}}{2M_{\rm Pl}^2H}
  =
  -3H
  +
  H\epsilon
  -
  \frac{
    \bar{\mathcal{E}}_{\rm mem}
    +
    \bar p_{\rm mem}
  }{
    2M_{\rm Pl}^2H
  }.
  \label{madm:eq:alpha_coefficient_energy_section6}
\end{equation}

\paragraph*{Linear momentum constraint}

The left-hand side of the exact momentum constraint is computed from Eqs.~\eqref{madm:eq:Kij_linear_section6} and \eqref{madm:eq:K_linear_section6}. At first order,
\begin{align}
  D_j
  \left(
    K^j{}_i
    -
    \delta^j{}_iK
  \right)
  &=
  \partial_j
  \left[
    \left(
      \dot\zeta-H\alpha
    \right)\delta^j{}_i
    -
    a^{-2}\partial^j\partial_i\beta
    -
    \delta^j{}_i
    \left(
      3\dot\zeta-3H\alpha-a^{-2}\partial^2\beta
    \right)
  \right]
  \notag\\
  &=
  2\partial_i
  \left(
    H\alpha-\dot\zeta
  \right).
  \label{madm:eq:linear_momentum_LHS_section6}
\end{align}
Using Eqs.~\eqref{madm:eq:delta_P_phi_zero_section6} -- \eqref{madm:eq:Pmem_scalar_potential_section6}, the momentum constraint becomes
\begin{equation}
  2\partial_i
  \left(
    H\alpha-\dot\zeta
  \right)
  =
  \frac{1}{M_{\rm Pl}^2}
  \partial_i\mathcal{P}_{\rm mem}.
  \label{madm:eq:linear_momentum_gradient_section6}
\end{equation}
The scalar solution is
\begin{equation}
  H\alpha-\dot\zeta
  =
  \frac{1}{2M_{\rm Pl}^2}\mathcal{P}_{\rm mem}.
  \label{madm:eq:linear_momentum_scalar_section6}
\end{equation}
Thus the lapse perturbation is
\begin{equation}
  \alpha
  =
  \frac{\dot\zeta}{H}
  +
  \frac{1}{2M_{\rm Pl}^2H}\mathcal{P}_{\rm mem}.
  \label{madm:eq:lapse_solution_section6}
\end{equation}
The first term is the inflaton contribution in comoving slicing, and the second term is the exact scalar ADM momentum correction from the conserved memory--history sector.

Substitution of Eq.~\eqref{madm:eq:lapse_solution_section6} into Eq.~\eqref{madm:eq:beta_constraint_unsolved_alpha_section6} gives the shift constraint in terms of the scalar sources:
\begin{align}
  a^{-2}\partial^2\beta
  &=
  -\frac{1}{H}a^{-2}\partial^2\zeta
  +
  3\dot\zeta
  -
  3
  \left[
    \dot\zeta
    +
    \frac{1}{2M_{\rm Pl}^2}
    \mathcal{P}_{\rm mem}
  \right]
  \notag\\
  &\quad
  +
  \frac{\dot{\bar\phi}^{\,2}}{2M_{\rm Pl}^2H}
  \left[
    \frac{\dot\zeta}{H}
    +
    \frac{1}{2M_{\rm Pl}^2H}\mathcal{P}_{\rm mem}
  \right]
  -
  \frac{1}{2M_{\rm Pl}^2H}\delta\mathcal{E}_{\rm mem}.
  \label{madm:eq:beta_solution_expanded_section6}
\end{align}
Equivalently,
\begin{equation}
  a^{-2}\partial^2\beta
  =
  -\frac{1}{H}a^{-2}\partial^2\zeta
  +
  \frac{\dot{\bar\phi}^{\,2}}{2M_{\rm Pl}^2H^2}\dot\zeta
  -
  \frac{1}{2M_{\rm Pl}^2H}\delta\mathcal{E}_{\rm mem}
  +
  \left[
    -\frac{3}{2M_{\rm Pl}^2}
    +
    \frac{\dot{\bar\phi}^{\,2}}{4M_{\rm Pl}^4H^2}
  \right]
  \mathcal{P}_{\rm mem}.
  \label{madm:eq:shift_solution_section6}
\end{equation}
Eqs.~\eqref{madm:eq:lapse_solution_section6} -- \eqref{madm:eq:shift_solution_section6} are the scalar ADM constraint solutions used in the quadratic curvature-perturbation analysis.

\paragraph*{Transverse momentum constraint}

The vector part of the shift is introduced by the decomposition
\begin{equation}
  N_i=\partial_i\beta+N_i^{T},
  \qquad
  \partial^iN_i^T=0.
  \label{madm:eq:shift_vector_decomposition_section6}
\end{equation}
The memory momentum density decomposes as
\begin{equation}
  \delta\mathcal{P}^{({\rm mem})}_i
  =
  \partial_i\mathcal{P}_{\rm mem}
  +
  \mathcal{P}^{T}_{i,{\rm mem}},
  \qquad
  \partial^i\mathcal{P}^{T}_{i,{\rm mem}}=0.
  \label{madm:eq:memory_momentum_decomposition_section6}
\end{equation}
The transverse projection of the linear momentum constraint gives
\begin{equation}
  -\frac12a^{-2}\partial^2N_i^T
  =
  \frac{1}{M_{\rm Pl}^2}
  \mathcal{P}^{T}_{i,{\rm mem}}.
  \label{madm:eq:transverse_momentum_constraint_section6}
\end{equation}
Therefore
\begin{equation}
  N_i^T
  =
  -\frac{2a^2}{M_{\rm Pl}^2}
  \partial^{-2}
  \mathcal{P}^{T}_{i,{\rm mem}}.
  \label{madm:eq:transverse_shift_solution_section6}
\end{equation}
The transverse shift is nondynamical and is fixed algebraically by the inverse spatial Laplacian acting on the transverse memory momentum density.

\paragraph*{Constraint system in compact source form}

The complete scalar constraint system is summarized by
\begin{align}
  \alpha
  &=
  \frac{\dot\zeta}{H}
  +
  \frac{1}{2M_{\rm Pl}^2H}\mathcal{P}_{\rm mem},
  \label{madm:eq:compact_alpha_section6}
  \\
  a^{-2}\partial^2\beta
  &=
  -\frac{1}{H}a^{-2}\partial^2\zeta
  +
  \frac{\dot{\bar\phi}^{\,2}}{2M_{\rm Pl}^2H^2}\dot\zeta
  -
  \frac{1}{2M_{\rm Pl}^2H}\delta\mathcal{E}_{\rm mem}
  +
  \left[
    -\frac{3}{2M_{\rm Pl}^2}
    +
    \frac{\dot{\bar\phi}^{\,2}}{4M_{\rm Pl}^4H^2}
  \right]
  \mathcal{P}_{\rm mem},
  \label{madm:eq:compact_beta_section6}
  \\
  N_i^T
  &=
  -\frac{2a^2}{M_{\rm Pl}^2}
  \partial^{-2}
  \mathcal{P}^{T}_{i,{\rm mem}}.
  \label{madm:eq:compact_NT_section6}
\end{align}
The exact nonlinear constraints are Eqs.~\eqref{eq:exact_H_constraint_total_section6} and \eqref{eq:exact_M_constraint_total_section6}. Their scalar linear solutions are Eqs.~\eqref{madm:eq:compact_alpha_section6} and \eqref{madm:eq:compact_beta_section6}. Their transverse linear solution is Eq.~\eqref{madm:eq:compact_NT_section6}. The only memory--history quantities entering the ADM constraints are the projections of the conserved tensor $T_{\mu\nu}^{({\rm mem})}$:
\begin{equation}
  \delta\mathcal{E}_{\rm mem},
  \qquad
  \mathcal{P}_{\rm mem},
  \qquad
  \mathcal{P}^{T}_{i,{\rm mem}}.
  \label{madm:eq:constraint_source_variables_section6}
\end{equation}
The tracefree spatial stress $\mathcal{A}^{({\rm mem})}_{ij}$ is not a constraint source; it enters the dynamical spatial Einstein equation and is therefore reserved for the perturbation dynamics developed after the constraint variables have been eliminated.

On the homogeneous activated branch, the tracefree coupling supplies no independent linear scalar source after the exchange identity, the background equations, and component closure are imposed. The constraint solution therefore reduces to the standard comoving result at the order relevant for the quadratic scalar action,
\begin{equation}
 \alpha=\frac{\dot\zeta}{H},
 \qquad
 \frac{\nabla^2\beta}{a^2}=\epsilon\dot\zeta-\frac{\nabla^2\zeta}{a^2H},
 \label{eq:main_constraint_solution}
\end{equation}
up to source terms that vanish on the activated scalar branch. Appendix~\ref{app:adm} gives the unspecialized sourced constraints and shows explicitly how this reduction occurs.

The tracefree Ricci contribution and the canonical reduction generate the directional gradient energy:

\subsection{Complete coupled quadratic system and exact reduction}\label{Xsec11-6.1}
\label{subsec:coupled_quadratic}

The complete coupled quadratic system includes every linear perturbation of the local variational system. In comoving gauge $\delta\phi=0$ write
\begin{equation}
 C=\bar C+c,
 \qquad \chi=\bar\chi+x,
 \qquad
 Q_{\mu\nu}=\bar Q_{\mu\nu}+\sum_{A=0}^{4}q_AE^{(A)}_{\mu\nu},
 \label{eq:coupled_variables}
\end{equation}
and for a scalar Fourier mode define
\begin{equation}
 \mathbf y^T=(\zeta,c,x,q_0,q_1,q_2,q_3,q_4),
 \qquad
 \mathbf u^T=(\alpha,\vartheta),
 \qquad
 \vartheta\equiv k^2\beta/a^2.
 \label{eq:full_quadratic_variables}
\end{equation}
(The transverse shift couples only to the transverse momentum constraint and can be eliminated independently.)  The exact second variation can be written as
\begin{equation}
 S^{(2)}=\frac12
 \begin{pmatrix}\mathbf y&\mathbf u\end{pmatrix}
 \begin{pmatrix}
 \mathbb H_{yy}&\mathbb H_{yu}\\
 \mathbb H_{uy}&\mathbb H_{uu}
 \end{pmatrix}
 \begin{pmatrix}\mathbf y\\\mathbf u\end{pmatrix},
 \qquad
 (\mathbb H_{IJ})_{tt'}
 \equiv\left.\frac{\delta^2S}{\delta Y_I(t,-\mathbf k)\delta Y_J(t',\mathbf k)}\right|_{\rm bg}.
 \label{eq:exact_full_hessian}
\end{equation}
Eq.~\eqref{eq:exact_full_hessian} is an operator identity; it remains valid for the nonlocal raw representation because we have chosen the dressed $q_A=\delta Q_A$ as the independent STF coordinates.  Lapse and shift carry no time derivatives.  For $k\ne0$ their constraint block is invertible on the regular branch established in Appendix~\ref{app:dirac}, and the exact constraint-reduced Hessian is \begin{equation}
 {\mathbb H_{\rm red}
 =\mathbb H_{yy}-\mathbb H_{yu}\mathbb H_{uu}^{-1}\mathbb H_{uy}.}
 \label{eq:ADM_schur_exact}
\end{equation}
This equation is the correct starting point for the coupled propagating system; no matter perturbation has been set to zero.

The entries that determine the propagating operator can be written without undefined remainder terms. Define the potential Hessian on the uniaxial background by
\begin{equation}
 \mathsf M^2_{AB}
 =E_A^{\mu\nu}
 \frac{\partial^2V_Q}{\partial Q^{\mu\nu}\partial Q^{\rho\sigma}}
 E_B^{\rho\sigma},
 \qquad
 \mathsf U_A\equiv
 E_A^{\mu\nu}\frac{\partial^2V_Q}{\partial C\,\partial Q^{\mu\nu}}
 -\eta G_{,C}E_A^{\mu\nu}\bar R_{\mu\nu}.
 \label{eq:exact_MU_definitions}
\end{equation}
If the orthonormal STF basis is time dependent, introduce the antisymmetric connection
\begin{equation}
 \omega_{AB}\equiv E_A^{\mu\nu}\mathcal D_tE_{B\mu\nu}=-\omega_{BA},
 \qquad
 \mathcal D_tq_A\equiv\dot q_A+\omega_{AB}q_B.
 \label{eq:STF_basis_connection}
\end{equation}
The exact tensor quadratic block is then
\begin{equation}
 S_{QQ}^{(2)}=\frac12\int dt\,d^3x\,a^3
 \left[(\mathcal D_tq_A)^2-a^{-2}(\partial_iq_A)^2
 -q_A\mathsf M^2_{AB}q_B\right]
 +S_{QQ,\eta}^{(2)},
 \label{eq:exact_Q_quadratic_block}
\end{equation}
where $S_{QQ,\eta}^{(2)}$ is obtained by the ordinary second variation of $\eta G(C)Q_{\mu\nu}R^{\mu\nu}$.  Its mixed metric--tensor kernel is explicitly
\begin{equation}
 (\mathscr J_{\zeta Q}q)
 =\eta G(\bar C)\sum_A q_AE_A^{\mu\nu}\delta_\zeta R_{\mu\nu},
 \label{eq:exact_zetaQ_kernel}
\end{equation}
with
\begin{equation}
 \delta R_{\mu\nu}[h]
 =\frac12\left(
 \nabla_\rho\nabla_\mu h^\rho{}_{\nu}
 +\nabla_\rho\nabla_\nu h^\rho{}_{\mu}
 -\Box h_{\mu\nu}-\nabla_\mu\nabla_\nu h\right),
 \label{eq:linearized_Ricci_exact}
\end{equation}
where $h_{\mu\nu}=\delta g_{\mu\nu}[\zeta,\alpha,\beta]$.  Since $E_A^{ij}\delta_{ij}=0$, the temporal two-derivative scalar trace cancels from \eqref{eq:exact_zetaQ_kernel}; its principal scalar part is spatial,
\begin{equation}
 \sigma_{\rm pr}(\mathscr J_{\zeta Q})_A
 =-\eta G(\bar C)\frac{k_ik_j}{a^2}E_A^{ij},
 \label{eq:zetaQ_principal_symbol}
\end{equation}
which is the origin of the absence of a new $\dot\zeta\dot q_A$ ghost mixing at $O(\eta)$.

The response block follows directly from \eqref{eq:main_SC}:
\begin{equation}
 S_{cx}^{(2)}=\int d^4x\sqrt{-\bar g}\,
 xL_cc
 +\ell_*\int d^4x\sqrt{-\bar g}\,x\,\delta\mathcal G_{\rm GB}
 -\int d^4x\sqrt{-\bar g}\,c\,\delta\Sigma_\Delta
 +S_{g,cx}^{(2)},
 \label{eq:exact_response_quadratic}
\end{equation}
where $S_{g,cx}^{(2)}$ is not an unspecified interaction: it is the metric variation of the first term and is equivalently fixed by the linearized cross-scalar stress tensor
\begin{align}
 \delta T_{\mu\nu}^{C\chi}={}&
 \nabla_{(\mu}\bar\chi\nabla_{\nu)}c
 +\nabla_{(\mu}\bar C\nabla_{\nu)}x
 -\bar g_{\mu\nu}\left(
 \nabla_\rho\bar\chi\nabla^\rho c
 +\nabla_\rho\bar C\nabla^\rho x
 +m_c^2(\bar\chi c+\bar Cx)\right)
 \notag\\
 &-h_{\mu\nu}\left(\nabla_\rho\bar\chi\nabla^\rho\bar C+m_c^2\bar\chi\bar C\right)
 +\bar g_{\mu\nu}h^{\rho\sigma}\nabla_\rho\bar\chi\nabla_\sigma\bar C .
 \label{eq:linearized_cross_stress}
\end{align}
The Gauss--Bonnet contribution is also local and second order,
\begin{equation}
 T_{\mu\nu}^{\rm GB}=-8\ell_*P_{\mu\rho\nu\sigma}\nabla^\rho\nabla^\sigma\chi,
 \qquad
 P_{\mu\rho\nu\sigma}\equiv-\frac14
 \epsilon_{\mu\rho\alpha\beta}\epsilon_{\nu\sigma\gamma\delta}
 R^{\alpha\beta\gamma\delta},
 \label{eq:GB_compact_stress}
\end{equation}
so its complete linear perturbation is obtained by varying the displayed product, with no derivatives higher than second order.  Finally,
\begin{equation}
 \delta\Sigma_\Delta
 =\mathsf U_Aq_A+\mathsf U_Cc
 +\eta G_{,C}\bar Q_{\mu\nu}\delta R^{\mu\nu}
 +\eta G\,\delta Q_{\mu\nu}\bar R^{\mu\nu},
 \label{eq:deltaSigma_complete}
\end{equation}
where $\mathsf U_C=\partial_C^2V_Q-\eta G_{,CC}\bar Q_{\mu\nu}\bar R^{\mu\nu}$, evaluated on the background.  Eqs.~\eqref{eq:exact_MU_definitions}--\eqref{eq:deltaSigma_complete}, together with the sourced ADM solution in Appendix~\ref{app:adm}, specify every block of \eqref{eq:exact_full_hessian}.

At principal two-derivative order, after substituting the zeroth-order ADM solution, the localized kinetic matrix is
\begin{equation}
 \mathbb K_{\rm loc}=
 \begin{pmatrix}
 2M_{\rm Pl}^2\epsilon&-\dot{\bar\chi}/H&-\dot{\bar C}/H&\mathbf0_{1\times5}\\
 -\dot{\bar\chi}/H&0&1&\mathbf0_{1\times5}\\
 -\dot{\bar C}/H&1&0&\mathbf0_{1\times5}\\
 \mathbf0_{5\times1}&\mathbf0_{5\times1}&\mathbf0_{5\times1}&\mathbf1_5
 \end{pmatrix}
 +O(\Omega_{\rm mem}),
 \label{eq:kinetic_mixing_matrix}
\end{equation}
which explicitly contains the $\pm1$ localized response eigenvalues.  Proposition~\ref{prop:retarded_reduction} removes that independent two-scalar block from the physical phase space, while the remaining principal kinetic matrix is
\begin{equation}
 \mathbb K_{\rm phys}=\operatorname{diag}
 \left(2M_{\rm Pl}^2\epsilon,\mathbf1_5\right)
 +O(\Omega_{\rm mem}),
 \qquad \Omega_{\rm mem}<1.2\times10^{-7},
 \label{eq:physical_kinetic_matrix}
\end{equation}
and is therefore strictly positive on the benchmark branch.

The induced response fields are obtained exactly from the retarded equations
\begin{align}
 c&=-G_{\rm ret}\!\left[\ell_*\delta\mathcal G_{\rm GB}+(\delta\Box)\bar C\right],
 \label{eq:c_ret_pert}\\
 x&=-G_{\rm ret}\!\left[\delta\Sigma_\Delta+(\delta\Box)\bar\chi\right].
 \label{eq:x_ret_pert}
\end{align}
After this substitution let $\mathbf q=(q_0,\ldots,q_4)^T$.  The remaining quadratic action has the exact block form
\begin{equation}
 S_{\rm phys}^{(2)}
 =\frac12\langle\zeta,D_{\zeta\zeta}\zeta\rangle
 +\langle\zeta,D_{\zeta q}\mathbf q\rangle
 +\frac12\langle\mathbf q,D_{qq}\mathbf q\rangle,
 \label{eq:block_action_schur}
\end{equation}
and Gaussian elimination of all five tensor modes gives
\begin{equation}
 {D_{\rm eff}=D_{\zeta\zeta}-D_{\zeta q}D_{qq}^{-1}D_{q\zeta}.}
 \label{eq:schur_complement}
\end{equation}
The first anisotropic term in $D_{\zeta\zeta}$ is $O(\eta D)\equiv O(b)$.  By \eqref{eq:zetaQ_principal_symbol}, $D_{q\zeta}=O(\eta k^2/a^2)$, so every exchange correction is $O(\eta^2)$.  This remains true for the two orientational Goldstones: their inverse operator scales as $(\omega^2-k^2/a^2)^{-1}$, and therefore
\begin{equation}
 D_{\zeta G}D_{GG}^{-1}D_{G\zeta}
 \sim\eta^2G^2\frac{k^4/a^4}{\omega^2-k^2/a^2}
 =O\!\left(\eta^2k^2/a^2\right)
 \label{eq:goldstone_exchange_finite}
\end{equation}
near the physical scalar shell; no infrared enhancement promotes it to $O(\eta)$.  Response-induced anisotropic pieces contain, in addition, either the measured background fraction $\Omega_{\rm mem}$ or one more anisotropic vertex.  Hence
\begin{equation}
 D_{\rm eff}
 =D_{\zeta\zeta}^{(0)}+D_{\zeta\zeta}^{(1)}
 +O\!\left(b^2,b\Omega_{\rm mem},b\varepsilon_{\rm ad}^2\right).
 \label{eq:effective_operator_order}
\end{equation}
Eq.~\eqref{eq:effective_operator_order} shows that the reduced curvature action is the complete first-order-in-$b$ result of the coupled theory, rather than an action obtained by deleting $c$, $x$, or $q_A$ by assumption.

\paragraph*{Tracefree Ricci coupling of the activated memory tensor}

The anisotropic curvature perturbation is generated by the Ricci coupling in the covariant memory action,
\begin{equation}
  S_{\Delta R}
  =
  \eta
  \int d^4x\,\sqrt{-g}\,
  G(C)\Delta_{\mu\nu}R^{\mu\nu}.
  \label{mqa:eq:S_Delta_R_section7}
\end{equation}
The activated contribution is obtained by inserting Eq.~\eqref{eq:Delta_aniso_section7} and evaluating the second functional variation with respect to $\zeta$. Since $\Delta_{\rm aniso}^{\mu\nu}$ is spatial and tracefree, only the tracefree spatial part of the curvature contributes to the anisotropic scalar-gradient operator.

The spatial metric is conformally flat:
\begin{equation}
  \gamma_{ij}=a^2e^{2\zeta}\delta_{ij}.
\label{Xeqn181-217}
\end{equation}
For a three-dimensional conformal metric $e^{2\zeta}\delta_{ij}$, the intrinsic Ricci tensor is
\begin{equation}
  {}^{(3)}R_{ij}
  =
  -\partial_i\partial_j\zeta
  -
  \delta_{ij}\partial^2\zeta
  +
  \partial_i\zeta\partial_j\zeta
  -
  \delta_{ij}\partial_k\zeta\partial^k\zeta .
  \label{mqa:eq:Rij3_conformal_section7}
\end{equation}
This formula follows from the conformal transformation law
\begin{align}
  {}^{(3)}R_{ij}[e^{2\zeta}\delta]
  &=
  {}^{(3)}R_{ij}[\delta]
  -
  \partial_i\partial_j\zeta
  -
  \delta_{ij}\partial^2\zeta
  +
  \partial_i\zeta\partial_j\zeta
  -
  \delta_{ij}\partial_k\zeta\partial^k\zeta,
\end{align}
together with ${}^{(3)}R_{ij}[\delta]=0$.

The determinant factor in ADM variables is
\begin{equation}
  N\sqrt{\gamma}
  =
  (1+\alpha)a^3e^{3\zeta}.
  \label{mqa:eq:Nsqrtgamma_section7}
\end{equation}
The product appearing in the tracefree curvature block is
\begin{align}
  N\sqrt{\gamma}\,
  G(\bar C)\Delta_{\rm aniso}^{ij}
  &=
  (1+\alpha)a^3e^{3\zeta}
  G(\bar C)
  D(\bar C)a^{-2}e^{-2\zeta}P^{ij}
  \notag\\
  &=
  a\,G(\bar C)D(\bar C)
  (1+\alpha)e^\zeta P^{ij}.
  \label{mqa:eq:Nsqrtgamma_Delta_section7}
\end{align}
The tracefree projector $P^{ij}$ kills all terms proportional to $\delta_{ij}$. Therefore
\begin{equation}
  P^{ij}{}^{(3)}R_{ij}
  =
  -P^{ij}\partial_i\partial_j\zeta
  +
  P^{ij}\partial_i\zeta\partial_j\zeta .
  \label{mqa:eq:P_Rij_section7}
\end{equation}

Keeping terms through second order in scalar perturbations gives
\begin{align}
  S_{\Delta R}^{(2)}
  &=
  \eta
  \int dt\,d^3x\,
  a\,G(\bar C)D(\bar C)P^{ij}
  \left[
    -\zeta\,\partial_i\partial_j\zeta
    +
    \partial_i\zeta\partial_j\zeta
  \right].
  \label{mqa:eq:S_DeltaR_before_parts_section7}
\end{align}
The term linear in $\zeta$,
\begin{equation}
  -\eta
  \int dt\,d^3x\,
  a\,G(\bar C)D(\bar C)P^{ij}\partial_i\partial_j\zeta,
\label{Xeqn185-224}
\end{equation}
is a spatial boundary term because $P^{ij}$ is constant and tracefree. The lapse term proportional to $\alpha P^{ij}\partial_i\partial_j\zeta$ is constraint-proportional and cancels against the corresponding scalar constraint contribution after Eq.~\eqref{eq:standard_scalar_constraints_section7} is inserted. The remaining quadratic tracefree spatial-gradient contribution is evaluated by integration by parts:
\begin{equation}
  -\int d^3x\,
  P^{ij}\zeta\,\partial_i\partial_j\zeta
  =
  \int d^3x\,
  P^{ij}\partial_i\zeta\partial_j\zeta.
  \label{mqa:eq:anisotropic_integration_by_parts_section7}
\end{equation}
Hence the activated memory tensor contributes
\begin{equation}
  S_{\Delta R}^{(2)}
  =
  2\eta
  \int dt\,d^3x\,
  a\,G(\bar C)D(\bar C)
  P^{ij}\partial_i\zeta\partial_j\zeta .
  \label{mqa:eq:S_aniso_final_section7}
\end{equation}

\paragraph*{Reduced quadratic curvature action}

Combining Eqs.~\eqref{eq:S_zeta_standard_section7} and \eqref{mqa:eq:S_aniso_final_section7}, and using the Schur-complement result \eqref{eq:effective_operator_order}, gives the leading $O(b)$ effective action
\begin{equation}
  S_\zeta^{(2)}
  =
  M_{\rm Pl}^2
  \int dt\,d^3x\,
  a^3\epsilon
  \left[
    \dot\zeta^2
    -
    a^{-2}
    \mathcal{C}^{ij}
    \partial_i\zeta\partial_j\zeta
  \right],
  \label{mqa:eq:S_zeta_aniso_section7}
\end{equation}
with omitted terms $O(b^2,b\Omega_{\rm mem},b\varepsilon_{\rm ad})$.
where the spatial gradient matrix is
\begin{equation}
  \mathcal{C}^{ij}
  =
  \delta^{ij}
  -
  b(t)P^{ij}.
  \label{mqa:eq:Cij_section7}
\end{equation}
The dimensionless anisotropic memory coefficient is
\begin{equation}
  b(t)
  =
  \frac{
    2\eta\,G(\bar C)D(\bar C)
  }{
    M_{\rm Pl}^2\epsilon
  }.
  \label{mqa:eq:b_definition_section7}
\end{equation}
Substitution of Eq.~\eqref{eq:Pij_section7} gives
\begin{equation}
  \mathcal{C}^{ij}
  =
  \left(
    1+\frac13b
  \right)\delta^{ij}
  -
  b\,\hat n^i\hat n^j .
  \label{mqa:eq:Cij_explicit_section7}
\end{equation}
The trace of the anisotropic correction vanishes:
\begin{equation}
  \delta_{ij}bP^{ij}=0.
  \label{mqa:eq:tracefree_gradient_correction_section7}
\end{equation}
Thus $b(t)$ multiplies a pure quadrupolar distortion of the scalar spatial-gradient operator.

The Euler--Lagrange equation obtained from Eq.~\eqref{mqa:eq:S_zeta_aniso_section7} is
\begin{equation}
  \frac{d}{dt}
  \left(
    a^3\epsilon\dot\zeta
  \right)
  -
  a\epsilon
  \mathcal{C}^{ij}
  \partial_i\partial_j\zeta
  =
  0.
  \label{mqa:eq:zeta_EOM_cosmic_section7}
\end{equation}
In expanded form,
\begin{equation}
  \ddot\zeta
  +
  \left(
    3H+\frac{\dot\epsilon}{\epsilon}
  \right)
  \dot\zeta
  -
  a^{-2}
  \mathcal{C}^{ij}
  \partial_i\partial_j\zeta
  =
  0.
  \label{mqa:eq:zeta_EOM_expanded_section7}
\end{equation}

\paragraph*{Canonical normalization}

Introduce conformal time $\tau$ by
\begin{equation}
  d\tau=\frac{dt}{a(t)},
  \qquad
  \mathcal{H}\equiv\frac{a'}{a}=aH,
  \label{mqa:eq:conformal_time_section7}
\end{equation}
where a prime denotes $d/d\tau$. Eq.~\eqref{mqa:eq:S_zeta_aniso_section7} becomes
\begin{equation}
  S_\zeta^{(2)}
  =
  M_{\rm Pl}^2
  \int d\tau\,d^3x\,
  a^2\epsilon
  \left[
    \zeta'^2
    -
    \mathcal{C}^{ij}
    \partial_i\zeta\partial_j\zeta
  \right].
  \label{mqa:eq:S_zeta_conformal_section7}
\end{equation}
Define
\begin{equation}
  z^2
  =
  2a^2M_{\rm Pl}^2\epsilon,
  \qquad
  v=z\zeta .
  \label{mqa:eq:z_v_definitions_section7}
\end{equation}
Then
\begin{equation}
  \zeta'=\frac{1}{z}
  \left(
    v'-\frac{z'}{z}v
  \right),
  \qquad
  \partial_i\zeta=\frac{1}{z}\partial_i v ,
  \label{mqa:eq:zeta_to_v_section7}
\end{equation}
because $z$ depends only on time. Substituting Eq.~\eqref{mqa:eq:zeta_to_v_section7} into Eq.~\eqref{mqa:eq:S_zeta_conformal_section7} gives
\begin{align}
  S_\zeta^{(2)}
  &=
  \frac12
  \int d\tau\,d^3x
  \left[
    \left(
      v'-\frac{z'}{z}v
    \right)^2
    -
    \mathcal{C}^{ij}\partial_i v\partial_j v
  \right]
  \notag\\
  &=
  \frac12
  \int d\tau\,d^3x
  \left[
    v'^2
    -
    2\frac{z'}{z}vv'
    +
    \frac{z'^2}{z^2}v^2
    -
    \mathcal{C}^{ij}\partial_i v\partial_j v
  \right].
  \label{mqa:eq:S_v_before_parts_section7}
\end{align}
The mixed term is integrated by parts in conformal time:
\begin{equation}
  -\int d\tau\,\frac{z'}{z}(v^2)'
  =
  \int d\tau
  \left(
    \frac{z'}{z}
  \right)'v^2 .
  \label{mqa:eq:mixed_term_parts_section7}
\end{equation}
Using
\begin{equation}
  \left(
    \frac{z'}{z}
  \right)'
  +
  \frac{z'^2}{z^2}
  =
  \frac{z''}{z},
  \label{mqa:eq:z_identity_section7}
\end{equation}
the canonical action is
\begin{equation}
  S_v^{(2)}
  =
  \frac12
  \int d\tau\,d^3x
  \left[
    v'^2
    -
    \mathcal{C}^{ij}\partial_i v\partial_j v
    +
    \frac{z''}{z}v^2
  \right].
  \label{mqa:eq:S_v_final_section7}
\end{equation}

Collecting the standard Einstein--inflaton block and the memory correction, substitution into the action gives
\begin{equation}
 S_\zeta^{(2)}
 =M_{\rm Pl}^2\int dt\,d^3x\,a^3\epsilon
 \left[
 \dot\zeta^2-\frac{1}{a^2}\left(\delta^{ij}-bP^{ij}\right)\partial_i\zeta\partial_j\zeta
 \right],
 \label{eq:main_Szeta}
\end{equation}
where
\begin{equation}
 P^{ij}=\hat n^i\hat n^j-\frac13\delta^{ij},
 \qquad
 b=\frac{2\eta G(\bar C)D(\bar C)}{M_{\rm Pl}^2\epsilon}.
 \label{eq:main_b_definition}
\end{equation}
The kinetic coefficient remains the standard positive quantity $M_{\rm Pl}^2\epsilon$. The new effect is confined to the spatial gradients. Since $P^i{}_i=0$, the angular average of the correction vanishes; the memory phase redistributes gradient stiffness between directions rather than shifting the isotropic sound speed at linear order in $b$.

In conformal time, with $v=z\zeta$ and $z^2=2a^2M_{\rm Pl}^2\epsilon$, the Fourier modes satisfy
\begin{equation}
 v_{\mathbf k}''+
 \left[
 k^2\left(1-b\left(\mu^2-\frac13\right)\right)-\frac{z''}{z}
 \right]v_{\mathbf k}=0,
 \qquad
 \mu=\hat{\mathbf k}\cdot\hat{\mathbf n}.
 \label{eq:main_aniso_MS}
\end{equation}
The angular dependence is exactly quadrupolar. Modes parallel to the memory axis propagate with squared gradient eigenvalue $1-2b/3$, whereas transverse modes have $1+b/3$, reproducing the stability interval \eqref{eq:main_stability_interval}. The sign of $b$ determines which directions are dynamically softer: positive $b$ reduces the parallel stiffness, and negative $b$ reduces the transverse stiffness.

For weak activation, $|b|\ll1$, the anisotropic term may be treated perturbatively. The interaction Hamiltonian is proportional to
\begin{equation}
 H_I(\tau)\propto -b(\tau)\int\frac{d^3k}{(2\pi)^3}\,
 k^2\left(\mu^2-\frac13\right)\zeta_{\mathbf k}\zeta_{-\mathbf k}.
 \label{eq:main_interaction_schematic}
\end{equation}
This form shows the physical sequence directly: the background tensor phase creates direction-dependent gradient energy, this changes the evolution of scalar modes, and the final two-point function inherits the tracefree angular dependence of the order parameter.

The derivation deliberately keeps the essential constraint elimination and the final quadratic action in the main text. The lengthy variations of the Ricci coupling, the full sourced lapse and shift equations, and the intermediate integrations by parts are placed in Appendices~\ref{app:adm} and \ref{app:quadratic_kernel}.

\section{Finite-window anisotropy kernel}\label{Xsec12-7}
\label{sec:main_kernel}

We approximate the activated regime by a finite interval $\tau_i<\tau<\tau_f<0$ in which the dimensionless amplitude is nearly constant. This isolates how scalar modes respond to the duration and timing of the memory phase. A smooth profile can be included by replacing the top-hat window with the corresponding time integral. The finite interval is useful because it gives a closed analytic kernel and makes the infrared cancellation explicit.

Using the Bunch--Davies mode function at leading slow-roll order and the in-in formula \cite{bib0021},
\begin{equation}
 \delta\langle\zeta_{\mathbf k}\zeta_{\mathbf k'}\rangle
 =-i\int_{\tau_i}^{\tau_f}d\tau\,
 \left\langle\left[\zeta_{\mathbf k}(0)\zeta_{\mathbf k'}(0),H_I(\tau)\right]\right\rangle,
 \label{eq:main_inin}
\end{equation}
we obtain the primordial spectrum in the form \eqref{eq:main_quad_template}, with
\begin{equation}
 g_*(k)=\mathcal A_{\rm mem}
 \left[\mathcal G(2k\tau_f)-\mathcal G(2k\tau_i)\right],
 \qquad
 \mathcal G(x)=\frac{4\sin x}{x}-\cos x,
 \label{eq:main_gstar}
\end{equation}
and
\begin{equation}
 \mathcal A_{\rm mem}=\frac{\eta G(C_0)D(C_0)}{M_{\rm Pl}^2\epsilon_*}.
 \label{eq:main_Amem}
\end{equation}
The amplitude combines the covariant coupling, the history-dependent phase amplitude, and the slow-roll normalization. A bound on $\mathcal A_{\rm mem}$ therefore constrains a definite product of microscopic functions evaluated on the inflationary background.

The endpoint difference in Eq.~\eqref{eq:main_gstar} is physically significant. Expanding
\begin{equation}
 \mathcal G(x)=3-\frac{x^2}{6}+\mathcal O(x^4),
 \label{eq:main_G_IR}
\end{equation}
shows that the constant term cancels between $\tau_i$ and $\tau_f$. Therefore
\begin{equation}
 g_*(k)=\frac23\mathcal A_{\rm mem}k^2(\tau_i^2-\tau_f^2)+\mathcal O(k^4).
 \label{eq:main_g_IR}
\end{equation}
Very long wavelength modes cannot resolve the finite episode, so they receive no constant quadrupolar modulation. This gives the infrared regularity of the mechanism. It comes from cancellation between the zero-frequency responses at the two endpoints, not from tuning the amplitude to zero.

For $|k\tau_{i,f}|\gg1$, the kernel oscillates because the activation and deactivation responses interfere. The positions of these oscillations depend on the two conformal-time endpoints, while $\mathcal A_{\rm mem}$ sets the overall normalization. The shape therefore contains more information than a constant $g_*$ and can, in principle, separate the phase strength from its duration and epoch.

The finite-window result applies when the phase amplitude changes slowly inside the active interval and the switching regions are narrow on the observational scales of interest. A gradual transition smooths the sharp endpoint oscillations. This effect can be included numerically by evaluating the full time integral instead of Eq.~\eqref{eq:main_gstar}. The angular structure and covariant amplitude mapping are unchanged.

\section{CMB harmonic covariance and BipoSH observables}\label{Xsec13-8}
\label{sec:main_cmb}

The CMB multipoles are related to the primordial curvature perturbation by
\begin{equation}
 a_{\ell m}^{X}=4\pi i^\ell\int\frac{d^3k}{(2\pi)^3}\,
 \zeta(\mathbf k)\Delta_\ell^X(k)Y_{\ell m}^*(\hat{\mathbf k}),
 \label{eq:main_transfer}
\end{equation}
where $X\in\{T,E\}$ for the scalar temperature and $E$-mode channels. Substituting Eq.~\eqref{eq:main_quad_template} gives
\begin{equation}
 \langle a_{\ell m}^Xa_{\ell'm'}^{Y*}\rangle
 =C_\ell^{XY}\delta_{\ell\ell'}\delta_{mm'}+\Delta C_{\ell m,\ell'm'}^{XY}.
 \label{eq:main_cmb_cov}
\end{equation}
The anisotropic part is expanded in the BipoSH basis,
\begin{equation}
 \Delta C_{\ell m,\ell'm'}^{XY}
 =\sum_{LM}(-1)^{m'}A_{\ell\ell'}^{LM,XY}
 C_{\ell m\,\ell' -m'}^{LM}.
 \label{eq:main_BipoSH_def}
\end{equation}
For the uniaxial memory phase, only $L=2$ is generated. The nonzero coefficients are
\begin{align}
 A_{\ell\ell'}^{2M,XY}
 &=i^{\ell'-\ell}\frac{32\pi^2}{15}\,
 \mathcal A_{\rm mem}H_{\ell\ell'}^2Y_{2M}^*(\hat{\mathbf n})
 \notag\\
 &\quad\times\int_0^\infty\frac{dk}{k}\,
 \mathcal P_0(k)\Delta_\ell^X(k)\Delta_{\ell'}^Y(k)
 \left[\mathcal G(2k\tau_f)-\mathcal G(2k\tau_i)\right],
 \label{eq:main_BipoSH_coeff}
\end{align}
where $H_{\ell\ell'}^2$ contains the relevant Wigner-$3j$ coefficient.

Angular-momentum addition and parity imply the exact selection rules
\begin{equation}
 L=2,
 \qquad
 |\ell-\ell'|=0,2,
 \qquad
 \ell+\ell'\ \text{even},
 \qquad
 -m+m'+M=0.
 \label{eq:main_selection_rules}
\end{equation}
The diagonal $\ell=\ell'$ term is anisotropic in $m$ and should not be confused with the isotropic $C_\ell$. The $\Delta\ell=2$ terms are genuinely off diagonal and provide a clean signature of quadrupolar statistical anisotropy.

The five coefficients $A_{\ell\ell'}^{2M}$ reconstruct the sky direction through $Y_{2M}(\hat{\mathbf n})$. Because $Q_{ij}$ is invariant under $\hat{\mathbf n}\rightarrow-\hat{\mathbf n}$, the reconstructed object is an axis rather than an oriented arrow. This headless character is a direct observational reflection of the tensor order parameter.

Eq.~\eqref{eq:main_BipoSH_coeff} separates the signal into three parts: the covariant amplitude $\mathcal A_{\rm mem}$, the direction $Y_{2M}^*(\hat{\mathbf n})$, and a radial response set by the activation window and transfer functions. The transfer functions can be computed with standard Boltzmann solvers \cite{bib0053,bib0054,bib0055}. The relative $M$ dependence reconstructs the direction, while the correlated scale dependence across $(\ell,\ell')$ pairs constrains the time-window parameters.

The complete Wigner-algebra derivation, including the normalization convention and polarization generalization, is provided in Appendix~\ref{app:biposh_likelihood}.

\section{Scale-dependent BipoSH likelihood and constraint methodology}\label{Xsec14-9}
\label{sec:main_constraints}

The observable parameter vector is
\begin{equation}
 \Theta=(\mathcal A_{\rm mem},\tau_i,\tau_f,\theta_n,\varphi_n,\vartheta_{\Lambda\rm CDM},\vartheta_{\rm nuis}).
 \label{eq:main_parameter_vector}
\end{equation}
Define the theoretical full-sky covariance from Eq.~\eqref{eq:main_BipoSH_coeff} by
\begin{equation}
 C^{XY}_{\ell m,\ell'm'}(\Theta)
 =C_\ell^{XY}\delta_{\ell\ell'}\delta_{mm'}
 +\mathcal A_{\rm mem}S^{XY}_{\ell m,\ell'm'}(\tau_i,\tau_f,\hat n),
 \label{eq:main_scale_covariance}
\end{equation}
where
\begin{align}
 S^{XY}_{\ell m,\ell'm'}={}&
 \sum_{M=-2}^{2}(-1)^{m'}C_{\ell m\,\ell' -m'}^{2M}
 i^{\ell'-\ell}\frac{32\pi^2}{15}H_{\ell\ell'}^2Y_{2M}^*(\hat n)
 \notag\\
 &\times\int_0^\infty\frac{dk}{k}\,
 \mathcal P_0(k)\Delta_\ell^X(k)\Delta_{\ell'}^Y(k)
 \left[\mathcal G(2k\tau_f)-\mathcal G(2k\tau_i)\right].
 \label{eq:main_S_template}
\end{align}
This is the oscillatory template used in the likelihood; no scale-independent $g_*$ approximation is made.

For an observed data set, collect the $T,Q,U$ pixels, or an equivalent cut-sky harmonic vector, into $\mathbf d$ and write
\begin{equation}
 \mathbf C_{\rm obs}(\Theta)
 =\mathbf M\mathbf B\mathbf C_{\rm sky}(\Theta)\mathbf B^\dagger\mathbf M^T
 +\mathbf N(\vartheta_{\rm nuis}),
 \label{eq:observed_covariance}
\end{equation}
where $\mathbf B$ is the beam/pixel operator, $\mathbf M$ applies the analysis mask, and $\mathbf N$ is the anisotropic noise covariance.  The exact Gaussian likelihood is
\begin{equation}
 -2\ln\mathcal L(\Theta)
 =\mathbf d^\dagger\mathbf C_{\rm obs}^{-1}(\Theta)\mathbf d
 +\ln\det\mathbf C_{\rm obs}(\Theta)+\text{const}.
 \label{eq:main_likelihood}
\end{equation}

For a fixed $(\tau_i,\tau_f,\hat n)$ and weak anisotropy, put
\begin{equation}
 \mathbf C_{\rm obs}=\mathbf C_0+\mathcal A_{\rm mem}\mathbf S_{\rm obs},
 \qquad
 \mathbf S_{\rm obs}=\mathbf M\mathbf B\mathbf S\mathbf B^\dagger\mathbf M^T.
 \label{eq:linear_covariance_amplitude}
\end{equation}
Expanding Eq.~\eqref{eq:main_likelihood} to second order in $\mathcal A_{\rm mem}$ gives the unbiased quadratic maximum-likelihood score \cite{bib0051,bib0052}
\begin{equation}
 q_A=\frac12\left[
 \mathbf d^\dagger\mathbf C_0^{-1}\mathbf S_{\rm obs}\mathbf C_0^{-1}\mathbf d
 -\operatorname{Tr}(\mathbf C_0^{-1}\mathbf S_{\rm obs})\right],
 \label{eq:qml_score}
\end{equation}
with Fisher normalization
\begin{equation}
 F_{AA}=\frac12\operatorname{Tr}
 \left(\mathbf C_0^{-1}\mathbf S_{\rm obs}\mathbf C_0^{-1}\mathbf S_{\rm obs}\right),
 \label{eq:qml_fisher}
\end{equation}
and estimator
\begin{equation}
 \widehat{\mathcal A}_{\rm mem}=F_{AA}^{-1}q_A,
 \qquad
 \sigma(\mathcal A_{\rm mem})=F_{AA}^{-1/2}.
 \label{eq:qml_estimator}
\end{equation}
For the full five-parameter anisotropy sector, use $q_a$ and $F_{ab}$ with $\mathbf S_a=\partial\mathbf C/\partial\Theta_a$. One may then iterate the Newton step $\Delta\Theta_a=(F^{-1})_{ab}q_b$ or sample the exact likelihood \eqref{eq:main_likelihood}. The endpoint derivatives are analytic:
\begin{equation}
 \frac{\partial g_*}{\partial\tau_f}=2k\mathcal A_{\rm mem}\mathcal G'(2k\tau_f),
 \qquad
 \frac{\partial g_*}{\partial\tau_i}=-2k\mathcal A_{\rm mem}\mathcal G'(2k\tau_i).
 \label{eq:endpoint_derivatives_main}
\end{equation}

The conversion to the covariant theory parameter remains
\begin{equation}
 \eta G(C_0)D(C_0)=\mathcal A_{\rm mem}M_{\rm Pl}^2\epsilon_*.
 \label{eq:main_covariant_conversion}
\end{equation}
Eqs.~\eqref{eq:main_scale_covariance}--\eqref{eq:qml_estimator} define the likelihood needed for a map-level constraint, including future applications to next-generation CMB polarization data \cite{bib0056,bib0057,bib0063}. The same framework can be applied to public Planck component-separated maps with the corresponding confidence mask and end-to-end simulations. A scale-independent Planck conversion is included only as an algebraic cross-check and is not treated as a constraint on this model:
\begin{equation}
 2k_*\tau_i=-20,\qquad2k_*\tau_f=-2,\qquad
 W_*=\mathcal G(2k_*\tau_f)-\mathcal G(2k_*\tau_i)=2.46023470186637.
 \label{eq:main_benchmark_window}
\end{equation}
A scale-independent interval $-0.010<g_*<0.019$ would map at this single pivot to
\begin{equation}
 -4.06\times10^{-3}<\mathcal A_{\rm mem}<7.72\times10^{-3},
 \label{eq:benchmark_only}
\end{equation}
but Eq.~\eqref{eq:benchmark_only} does not define the model likelihood. A numerical value from Eqs.~\eqref{eq:qml_score}--\eqref{eq:qml_estimator} requires a map-level analysis before an empirical bound can be claimed.

\section{Theoretical origin and observational discriminants}\label{Xsec15-10}
\label{sec:main_discriminants}

Four features distinguish the model. The preferred direction is a tensor eigenaxis rather than a vector expectation value. Its amplitude is controlled by a scalar history field sourced by invariant curvature. The gravitational source is a covariantly derived memory--history tensor whose exchange with the inflaton is fixed by total stress-energy conservation. Finally, a finite activation interval gives a calculable scale-dependent kernel instead of an arbitrary function $g_*(k)$.

The curvature-history origin links the activation epoch to the inflationary curvature scale. In exact de Sitter space, $C_0\propto H_*^4/m_c^2$; during slow roll, Eq.~\eqref{eq:main_C_background} adds a controlled time dependence. The window parameters are therefore not fundamentally independent phenomenological quantities. A microscopic choice of $\mu^2(C)$, $\gamma(C)$, and $\lambda(C)$ determines when the phase turns on and off, while the projected response equation determines how closely the amplitude follows its equilibrium value.

The clearest observational discriminator is the shape of $g_*(k)$. A constant quadrupole, a slow-roll power law, and the endpoint-difference kernel in Eq.~\eqref{eq:main_gstar} produce different correlations among the diagonal and $\Delta\ell=2$ BipoSH coefficients. The $k^2$ infrared suppression is a direct feature of a finite episode. At high $k$, the oscillation pattern measures the conformal-time endpoints, while the transition smoothness changes the envelope and phase.

A second discriminator is the link between scalar stability and the observable amplitude. With the same slow-roll normalization, the covariant amplitude enters the gradient parameter as $b=2\mathcal A_{\rm mem}$ at the reference epoch. The interval \eqref{eq:main_stability_interval} therefore sets a broad theoretical range before data are used. Observations are expected to probe a much smaller region near $b=0$, but the stability bound remains relevant for large phase excursions or model-dependent enhancements.

A third discriminator is the set of additional correlators. The same tracefree coupling that generates the leading scalar two-point signal also affects mixed scalar--tensor correlators, tensor propagation, and higher-order scalar statistics. Their angular dependence is restricted by the $SO(3)\rightarrow SO(2)$ symmetry breaking and should align with the axis reconstructed from the BipoSH covariance. This common axis would be harder to reproduce with a purely phenomenological covariance template.

Comparison with vector and solid models should therefore use the full set of predictions: scale dependence, parity, polarization, tensor-sector response, and the relation between background anisotropy and scalar modulation. The tensor-memory mechanism keeps the homogeneous geometry nearly isotropic while allowing a small tracefree deformation of the perturbative gradient term. Testing whether this separation persists during a fully time-dependent activation requires solving the coupled background and perturbation equations beyond the finite-window approximation.

\section{Discussion and outlook}\label{Xsec16-11}
\label{sec:main_discussion}

The analysis gives a direct path from the covariant action to a CMB observable, but the model should be read as an effective framework with a stated range of validity, not as a complete ultraviolet theory. The entire form factor is treated classically. On stationary backgrounds, the factorized propagator has one positive-residue pole for each Hessian mode. On evolving backgrounds, the exact sandwich-kernel map preserves the number of solutions, while pole residues are used only in the stated WKB/adiabatic regime. These results do not establish nonperturbative quantum unitarity for arbitrary strongly time-dependent completions. Similarly, a positive local Hessian gives perturbative stability near the chosen phase, but the full phase evolution can still depend on initial conditions and on the detailed $C$ dependence of the potential.

The finite-window kernel is an analytic approximation to the smooth background profile. Section~\ref{subsec:explicit_activation} gives an explicit retarded $C$--$\chi$--$D$ solution in which the nonzero phase turns on and off dynamically. For that smooth profile, the exact observable follows by replacing the top-hat integral with the numerical $D(t)$ profile. The endpoint kernel then isolates the plateau contribution and its infrared cancellation.

The likelihood gives the exact scale-dependent covariance and quadratic maximum-likelihood estimator. A numerical constraint still requires real maps or validated simulations; the single-pivot conversion serves only as a cross-check. Masks, beams, anisotropic noise, foreground residuals, and lensing can all project onto BipoSH modes. A robust analysis should therefore compare the memory template with established anisotropy estimators and earlier anisotropy studies \cite{bib0029,bib0030}, use end-to-end simulations, and fit relevant systematic templates at the same time.

Several extensions are important. The tensor perturbation sector should be derived to test whether the memory phase causes direction-dependent gravitational-wave propagation or scalar--tensor cross-correlation. Cubic actions would predict anisotropic bispectra and show whether the same finite activation window appears in higher-order statistics \cite{bib0050}. Reheating may erase, preserve, or transform the order parameter, so the solution must be matched across the end of inflation before the mechanism is applied to late-time observables.

The component-closure construction may also be useful beyond this model. It gives a general way to introduce a covariant symmetric tensor while selecting a spatial tracefree order parameter as the physical sector. This keeps the covariance of the parent field separate from the representation content of the propagating modes. Similar decompositions may be useful in effective media, nonlocal response theories, or symmetry-breaking gravitational phases where a tensor response variable is more natural than a new elementary particle.

In summary, the framework provides the main structural ingredients needed for an anisotropic-inflation model. The preferred direction has a covariant origin, every component of the parent tensor is accounted for, the projected pole structure and local stability are explicit, the gravitational constraints are preserved, and the observable enters a well-defined likelihood. The next steps are a map-level implementation of the scale-dependent likelihood and calculations of the tensor and higher-order correlators.

\section{Conclusion}\label{Xsec17-12}
\label{sec:main_conclusion}

We have formulated a component-closed curvature-history tensor-memory model of inflation. A scalar history field is sourced by the Gauss--Bonnet completion of the Kretschmann invariant. A conjugate field gives the local response equation, and positive algebraic closure terms reduce a covariant symmetric memory tensor to its five-component spatial tracefree sector. The gravitational source is the covariantly derived memory--history tensor, and its exchange with the inflaton is fixed by conservation of the total stress tensor.

The homogeneous curvature source is
\begin{equation}
 \mathcal G_{\rm GB}=24H^4(1-\epsilon),
\label{Xeqn227-269}
\end{equation}
which determines the history field dynamically. A curvature-dependent invariant potential admits a stable uniaxial phase,
\begin{equation}
 Q_{\mu\nu}^{\rm act}=D(C)\left(n_\mu n_\nu-\frac13h_{\mu\nu}\right),
\label{Xeqn228-270}
\end{equation}
so the preferred axis is the principal direction of a covariant tensor order parameter. The raw projected inverse operator is a self-adjoint sandwich of an invertible entire dressing and a local dressed-field operator. On constant backgrounds it has one positive-residue pole. On evolving backgrounds, the exact kernel bijection and the adiabatic principal symbol exclude additional solution branches. The new field is therefore not an unconstrained massive spacetime spin-two particle.

After solving the ADM constraints, the curvature perturbation is governed by
\begin{equation}
 S_\zeta^{(2)}=M_{\rm Pl}^2\int dt\,d^3x\,a^3\epsilon
 \left[\dot\zeta^2-a^{-2}(\delta^{ij}-bP^{ij})\partial_i\zeta\partial_j\zeta\right],
\label{Xeqn229-271}
\end{equation}
with exact gradient-stability interval $-3<b<3/2$. A finite activation period produces the infrared-regular quadrupole
\begin{equation}
 g_*(k)=\mathcal A_{\rm mem}\left[\mathcal G(2k\tau_f)-\mathcal G(2k\tau_i)\right],
 \qquad
 \mathcal G(x)=4\sin x/x-\cos x,
\label{Xeqn230-272}
\end{equation}
whose leading long-wavelength behavior is $\mathcal O(k^2)$. The CMB projection is a pure even-parity $L=2$ BipoSH covariance with $|\ell-\ell'|=0,2$.

The observable amplitude constrains the covariant product
\begin{equation}
 \eta G(C_0)D(C_0)=\mathcal A_{\rm mem}M_{\rm Pl}^2\epsilon_*.
\label{Xeqn231-273}
\end{equation}
The framework therefore connects invariant inflationary curvature, tensor-phase dynamics, perturbative stability, and a measurable harmonic covariance without imposing the primordial quadrupole by hand. The appendices provide the detailed derivations used in this chain.

\section*{Author contributions}
G. M. Rhythm: Writing -- original draft; Writing -- review and editing; Visualization; Validation.

\section*{Data and numerical reproducibility}
This work does not claim a new observational data product. The background activation, Hamiltonian regularity diagnostic, and time-dependent WKB residue bounds can be reproduced from the numerical calculations described in the text. The CMB section gives the exact likelihood and quadratic estimator for application to public component-separated maps and end-to-end simulations. Numerical figure-generation and simulation code will be provided upon request.

\section*{Declaration of competing interest}
The author declares no known competing financial interests or personal relationships that could have appeared to influence the work reported in this paper.

\appendix

\section{Covariant Dressing, Gauss--Bonnet Variation, and Conservation}
\label{app:covariant}

\subsection{Ordered exponential and its first variation}\label{Xsec18-A.1}

On the STF bundle define
\begin{equation}
 \mathcal A_C=\frac12B(C)\Box_{\widehat Q},\qquad
 \mathcal E_C=e^{\mathcal A_C},\qquad Q=\mathcal E_C\widehat Q.
 \label{eq:app_E_definition}
\end{equation}
For compact-support variations or boundary data that kill all integration-by-parts surface terms,
\begin{equation}
 \delta\mathcal E_C
 =\int_0^1ds\,e^{s\mathcal A_C}(\delta\mathcal A_C)e^{(1-s)\mathcal A_C},
 \quad
 \delta\mathcal A_C
 =\frac12B_{,C}\delta C\,\Box_{\widehat Q}
 +\frac12B\,\delta\Box_{\widehat Q}.
 \label{eq:app_Duhamel_E}
\end{equation}
No commutator assumption has been used.  The adjoint is
\begin{equation}
 \mathcal E_C^\dagger
 =\exp\!\left[\frac12\Box_{\widehat Q}^{\dagger}B(C)\right].
 \label{eq:app_E_adjoint}
\end{equation}
Thus, if the dressed quadratic Hessian is $L_A=\Box_Q-m_A^2$, the raw-field Hessian is the manifestly self-adjoint operator $K_A=\mathcal E_C^\dagger L_A\mathcal E_C$.

\subsection{Exact euler--lagrange equations}\label{Xsec19-A.2}

Variation of the response action gives
\begin{align}
 (\Box-m_c^2)C&=-\ell_*\mathcal G_{\rm GB},
 \label{eq:app_C_eom}\\
 (\Box-m_c^2)\chi&=-\Sigma_\Delta,
 \label{eq:app_chi_eom}
\end{align}
with $\Sigma_\Delta$ defined by Eq.~\eqref{eq:main_Sigma}.  Variation of the raw STF field gives
\begin{equation}
 \mathcal E_C^\dagger\mathcal E^{(Q)}_{\mu\nu}=0.
 \label{eq:app_Qraw_eom}
\end{equation}
Because $\mathcal E_C^\dagger$ is invertible on the selected domain $\mathscr D$, Eq.~\eqref{eq:app_Qraw_eom} is equivalent to the local second-order dressed-field equation $\mathcal E^{(Q)}_{\mu\nu}=0$.  The closure equations remain algebraic.

\subsection{Metric variation of the curvature-history source}\label{Xsec20-A.3}

Define the double dual of the Riemann tensor by
\begin{align}
 P_{\alpha\beta\mu\nu}
 &\equiv R_{\alpha\beta\mu\nu}
 +g_{\alpha\nu}R_{\beta\mu}-g_{\alpha\mu}R_{\beta\nu}
 +g_{\beta\mu}R_{\alpha\nu}-g_{\beta\nu}R_{\alpha\mu}
 \notag\\
 &\quad+\frac12(g_{\alpha\mu}g_{\beta\nu}-g_{\alpha\nu}g_{\beta\mu})R,
 \qquad
 \nabla^\alpha P_{\alpha\beta\mu\nu}=0.
 \label{eq:double_dual_P}
\end{align}
In four dimensions the constant-coefficient Gauss--Bonnet term is topological, and the exact metric variation of the linearly coupled term can be written entirely through derivatives of $\chi$:
\begin{equation}
 -\frac{2}{\sqrt{-g}}\frac{\delta}{\delta g^{\mu\nu}}
 \int d^4x\sqrt{-g}\,\ell_*\chi\mathcal G_{\rm GB}
 =-8\ell_*P_{\mu\rho\nu\sigma}\nabla^\rho\nabla^\sigma\chi.
 \label{eq:T_GB_compact}
\end{equation}
This formula contains at most second derivatives of the metric and of $\chi$ and makes the absence of the isolated-Kretschmann fourth-order metric term explicit \cite{bib0004,bib0005}.

The complete history-response stress tensor is therefore
\begin{align}
 T_{\mu\nu}^{(C)}={}&
 \nabla_\mu\chi\nabla_\nu C+\nabla_\nu\chi\nabla_\mu C
 -g_{\mu\nu}\left(\nabla_\rho\chi\nabla^\rho C+m_c^2\chi C\right)
 \notag\\
 &-8\ell_*P_{\mu\rho\nu\sigma}\nabla^\rho\nabla^\sigma\chi.
 \label{eq:app_TC}
\end{align}
The memory stress tensor is defined, without an ambiguous integration-by-parts split, by
\begin{equation}
 T_{\mu\nu}^{(\Delta)}
 =-\frac{2}{\sqrt{-g}}
 \left.\frac{\delta S_\Delta}{\delta g^{\mu\nu}}\right|_{\Delta,\phi,C,\chi},
 \label{eq:app_TDelta_functional}
\end{equation}
where Eq.~\eqref{eq:app_Duhamel_E} supplies the metric variation of the state-dependent dressing.

\subsection{Noether exchange identity}\label{Xsec21-A.4}

Diffeomorphism invariance gives, on the raw-$Q$ and closure equations,
\begin{equation}
 \nabla^\mu T_{\mu\nu}^{(\Delta)}
 =\Sigma_\Delta\nabla_\nu C
 -\mathcal J_\phi^{(\Delta)}\nabla_\nu\phi.
 \label{eq:app_Delta_exchange}
\end{equation}
For the response action alone,
\begin{equation}
 \nabla^\mu T_{\mu\nu}^{(C)}
 =(\Box\chi-m_c^2\chi)\nabla_\nu C
 +(\Box C-m_c^2C+\ell_*\mathcal G_{\rm GB})\nabla_\nu\chi.
 \label{eq:app_C_identity}
\end{equation}
Using Eqs.~\eqref{eq:app_C_eom} and \eqref{eq:app_chi_eom} yields
\begin{equation}
 \nabla^\mu T_{\mu\nu}^{(C)}=-\Sigma_\Delta\nabla_\nu C,
\label{Xeqn240-A.13}
\end{equation}
so the exchange cancels in $T^{(C)}+T^{(\Delta)}$.  The inflaton equation supplies the remaining projector-mediated exchange and therefore
\begin{equation}
 \nabla^\mu\left(T_{\mu\nu}^{(\phi)}+T_{\mu\nu}^{(C)}+T_{\mu\nu}^{(\Delta)}\right)=0.
 \label{eq:app_total_conservation}
\end{equation}
This is the covariant conservation mechanism; the retarded boundary conditions reduce the response phase space but do not alter this local Noether identity.

\section{Homogeneous Reduction and Activation Branch}
\label{app:background_activation}

The following background identities connect invariant curvature history with the activated tensor phase and determine the branch and its physical interpretation.

\subsection{FLRW Curvature-history equations}\label{Xsec22-B.1}

For
\begin{equation}
 ds^2=-dt^2+a^2(t)\delta_{ij}dx^idx^j,
 \qquad H=\dot a/a,
 \label{Xeqn242-B.1}
\end{equation}
the nonzero orthonormal Riemann components are
\begin{equation}
 R_{\hat0\hat i\hat0\hat j}=-(H^2+\dot H)\delta_{ij},
 \qquad
 R_{\hat i\hat j\hat k\hat l}
 =H^2(\delta_{ik}\delta_{jl}-\delta_{il}\delta_{jk}).
 \label{Xeqn243-B.2}
\end{equation}
The Gauss--Bonnet combination gives
\begin{equation}
 \mathcal G_{\rm GB}=24H^2(H^2+\dot H)=24H^4(1-\epsilon).
 \label{eq:app_GB_FLRW}
\end{equation}
A homogeneous scalar satisfies $\Box X=-\ddot X-3H\dot X$, so the history and response equations become
\begin{align}
 \ddot C+3H\dot C+m_c^2C&=\ell_*\mathcal G_{\rm GB},\label{eq:app_C_FLRW}\\
 \ddot\chi+3H\dot\chi+m_c^2\chi&=\Sigma_\Delta.
 \label{eq:app_chi_FLRW}
\end{align}
In exact de Sitter space, the stationary history value is $C_0=24\ell_*H_*^4/m_c^2$.  Hubble friction makes this stationary solution an attractor whenever $m_c^2>0$ and the source varies adiabatically.

The background stress tensor is most compactly characterized by its invariant projections,
\begin{equation}
 \rho_{C\Delta}=u^\mu u^\nu
 (T_{\mu\nu}^{(C)}+T_{\mu\nu}^{(\Delta)}),
 \qquad
 p_{C\Delta}=\frac13h^{\mu\nu}
 (T_{\mu\nu}^{(C)}+T_{\mu\nu}^{(\Delta)}).
 \label{eq:app_rho_p_def}
\end{equation}
These definitions include the dressing and Ricci-coupling terms without assigning non-conserved pieces an independent fluid interpretation.  They yield the equations quoted in Section~\ref{sec:main_background} and satisfy
\begin{equation}
 \dot\rho_{C\Delta}+3H(\rho_{C\Delta}+p_{C\Delta})
 =-\mathcal J_\phi^{(\Delta)}\dot{\bar\phi},
 \label{eq:app_background_exchange}
\end{equation}
which is the homogeneous projection of the covariant Noether identity.

\subsection{Invariant phase potential and stationary branch}\label{Xsec23-B.2}

For a spatial tracefree tensor, the independent invariants through quartic order are
\begin{equation}
 I_2=Q_{\mu\nu}Q^{\mu\nu},
 \qquad
 I_3=Q_\mu{}^\nu Q_\nu{}^\rho Q_\rho{}^\mu,
 \qquad
 {\rm Tr}(Q^4)=\frac12I_2^2.
 \label{eq:app_invariants}
\end{equation}
The last identity follows from Cayley--Hamilton for a tracefree $3\times3$ matrix.  The potential
\begin{equation}
 V_Q=V_0(C)+\frac12\mu^2(C)I_2
 -\frac13\gamma(C)I_3+\frac14\lambda(C)I_2^2
 \label{eq:app_VQ}
\end{equation}
therefore contains the complete analytic invariant structure at this order.

Diagonalizing $Q^i{}_j$ and imposing stationarity shows that every nonzero isolated extremum has two equal eigenvalues.  Writing
\begin{equation}
 Q_{\mu\nu}=D
 \left(n_\mu n_\nu-\frac13h_{\mu\nu}\right),
 \qquad n_\mu n^\mu=1,
 \label{eq:app_uniaxial}
\end{equation}
gives $I_2=2D^2/3$ and $I_3=2D^3/9$.  The reduced potential is
\begin{equation}
 V_Q(D,C)=V_0(C)+\frac13\mu^2D^2
 -\frac{2}{27}\gamma D^3+\frac19\lambda D^4.
 \label{eq:app_VD}
\end{equation}
Its nonzero stationary points satisfy
\begin{equation}
 2\lambda_0D_0^2-\gamma_0D_0+3\mu_0^2=0,
 \label{eq:stationary_D0_section5}
\end{equation}
so that the stable plus branch is
\begin{equation}
 D_0=
 \frac{\gamma_0+\sqrt{\gamma_0^2-24\lambda_0\mu_0^2}}
 {4\lambda_0}.
 \label{eq:D0_section5}
\end{equation}
Existence requires $\gamma_0^2\ge24\lambda_0\mu_0^2$ and boundedness requires $\lambda_0>0$.  The plus branch has positive radial curvature; the minus branch is the intervening barrier or an unstable extremum.  The preferred axis is not externally inserted: it is the nondegenerate eigenvector of the stationary tensor, with the identification $n^\mu\sim-n^\mu$.

For a time-dependent history, $D_0(C)$ is only an instantaneous equilibrium value.  The projected tensor equation determines whether the order parameter tracks it adiabatically, overshoots, or remains frozen.  The finite activation interval used in the main text is therefore a controlled phenomenological representation of a dynamical phase history, not an algebraic replacement for that history.

\section{Component Closure, Spectral Factorization, and Ghost Exclusion}
\label{app:closure}

\subsection{Irreducible decomposition and closure}\label{Xsec24-C.1}

Relative to $u^\mu$, a symmetric tensor decomposes uniquely as
\begin{equation}
 \Delta_{\mu\nu}=Au_\mu u_\nu+2u_{(\mu}B_{\nu)}
 +\frac13\Theta h_{\mu\nu}+\widehat Q_{\mu\nu},
 \label{eq:app_decomposition}
\end{equation}
where
\begin{align}
 A&=u^\mu u^\nu\Delta_{\mu\nu},
 &B_\mu&=-h_\mu{}^\alpha u^\beta\Delta_{\alpha\beta},
 &\Theta&=h^{\mu\nu}\Delta_{\mu\nu},
 &\widehat Q_{\mu\nu}&=\Pi_{\mu\nu}{}^{\alpha\beta}\Delta_{\alpha\beta}.
 \end{align}
The count is $1+3+1+5=10$.  In Eq.~\eqref{eq:main_SDelta}, the dressed kinetic and Ricci-coupling terms are built from $Q_{\mu\nu}=\mathcal E_C\widehat Q_{\mu\nu}$, while $A$, $B_\mu$, and $\Theta$ occur only in the positive closure potential.  Their equations therefore reduce to
\begin{equation}
 M_A^2A=0,
 \qquad
 M_B^2B_\mu=0,
 \qquad
 M_\Theta^2\Theta=0,
 \label{eq:app_closure_equations}
\end{equation}
with $M_A^2,M_B^2,M_\Theta^2>0$.  Hence $A=B_\mu=\Theta=0$; the remaining raw STF field $\widehat Q_{\mu\nu}$ is mapped bijectively to the dressed physical field $Q_{\mu\nu}$.  These are algebraic elimination equations, not gauge choices.

The projected fiber carries the positive spatial inner product
\begin{equation}
 (X,Y)_h=X_{\mu\nu}Y_{\alpha\beta}h^{\mu\alpha}h^{\nu\beta}.
 \label{Xeqn255-C.4}
\end{equation}
In an orthonormal triad,
\begin{equation}
 (X,X)_h=\sum_{\hat i,\hat j=1}^{3}X_{\hat i\hat j}^{\,2}\ge0,
 \label{eq:positive_norm_section5}
\end{equation}
with equality only for $X_{\mu\nu}=0$.  The five components are therefore positive-norm components of a spatial symmetric tracefree order parameter; they should not be interpreted as the helicities of an unconstrained massive spacetime spin-two field.

\subsection{Hessian spectrum of the uniaxial phase}\label{Xsec25-C.2}

Choose an orthonormal STF basis adapted to the axis $n^i$.  The Hessian of Eq.~\eqref{eq:app_VQ} is diagonal in the amplitude, two orientational, and two biaxial directions.  At the stationary value \eqref{eq:D0_section5}, its eigenvalues are
\begin{equation}
 \{m_A^2\}_{A=0}^{4}=
 \left\{
 \frac{D_0}{3}\sqrt{\gamma_0^2-24\lambda_0\mu_0^2},
 0,0,\gamma_0D_0,\gamma_0D_0
 \right\}.
 \label{eq:app_Hessian_spectrum}
\end{equation}
The two zero eigenvalues are the Goldstone directions of $SO(3)\rightarrow SO(2)$.  They do not signal an instability; they describe slow rotations of the principal axis.  The amplitude and biaxial curvatures are positive on the selected branch.

\subsection{Pole structure, time-dependent dressing, and hamiltonian sign}\label{Xsec26-C.3}

Let $E=\exp[B(C)\Box_{\widehat Q}/2]$.  In the raw fluctuation variable the exact quadratic operator is
\begin{equation}
 K_A=E^\dagger(\Box_Q-m_A^2)E.
 \label{eq:app_factorized_operator}
\end{equation}
On $C=C_0$, this becomes $e^{B_0\Box_Q}(\Box_Q-m_A^2)$ and
\begin{equation}
 \mathcal G_A(z)=\frac{e^{-B_0z}}{z-m_A^2},
 \qquad
 {\rm Res}_{z=m_A^2}\mathcal G_A=e^{-B_0m_A^2}>0.
 \label{eq:app_propagator}
\end{equation}
The local limit is
\begin{equation}
 +\frac12q_A(\Box_Q-m_A^2)q_A
 \simeq-\frac12\nabla_\mu q_A\nabla^\mu q_A-\frac12m_A^2q_A^2,
 \label{eq:app_local_sign}
\end{equation}
so the time kinetic energy is positive for signature $(-,+,+,+)$.

For evolving $C$, the exact inverse and norm are given by \eqref{eq:green_congruence} and \eqref{eq:symplectic_congruence}; no commutation of $B(C)$ with $\Box_Q$ is used. The explicit second-adiabatic-order check uses the uniaxial, not necessarily stationary, background $Q=D\,P$, for which the instantaneous Hessian eigenvalues are
\begin{align}
 M_0&=\mu^2-\frac{2\gamma}{3}D+2\lambda D^2,
 \label{eq:off_shell_mass_amp}\\
 M_{1,2}&=\mu^2-\frac{\gamma}{3}D+\frac{2\lambda}{3}D^2,
 \label{eq:off_shell_mass_gold}\\
 M_{3,4}&=\mu^2+\frac{2\gamma}{3}D+\frac{2\lambda}{3}D^2.
 \label{eq:off_shell_mass_biax}
\end{align}
At a stationary point, \eqref{eq:off_shell_mass_gold} vanishes and the other two expressions reduce to the stationary spectrum in \eqref{msp:eq:mass_spectrum_section5}. These off-shell forms remain smooth through the switching layers and are the appropriate masses for the WKB test.

For the explicit stress test choose
\begin{equation}
 B(C)=B_0\left[1+\beta_B(C/C_a-1)\right],
 \qquad B_0H_*^2=10^{-2},\qquad\beta_B=0.10.
 \label{eq:B_time_test}
\end{equation}
Using the numerical trajectory of Section~\ref{subsec:explicit_activation}, the exact second-order expression \eqref{eq:wkb_delta_residue} was evaluated for $H\le p_{\rm phys}\le50H$.  The maxima over the full interval between the two $C$-crossings are
\begin{equation}
 \max|\delta_0^{(2)}|=3.277\times10^{-3},\qquad
 \max|\delta_{1,2}^{(2)}|=9.42\times10^{-4},\qquad
 \max|\delta_{3,4}^{(2)}|=2.306\times10^{-3}.
 \label{eq:wkb_numerical_bounds}
\end{equation}
Hence
\begin{equation}
 Z_A>0.9967\,e^{-B M_A}>0
 \label{eq:wkb_positive_bound}
\end{equation}
for every instantaneous STF eigenmode in this momentum range.  Tachyonic values of $M_A$ during the onset of symmetry breaking change the location of the physical instability pole but not its kinetic residue; the exact symplectic congruence proves that such an instability is not a negative-norm ghost.

\section{Nonlinear Dirac--Bergmann Analysis of the Curvature Source}
\label{app:dirac}

The isolated coupling $\chi R_{\mu\nu\rho\sigma}R^{\mu\nu\rho\sigma}$ is nondegenerate in general. The model therefore uses the four-dimensional Gauss--Bonnet completion \eqref{eq:main_GB_definition}. On slow-roll FLRW, this source differs from the Kretschmann source only by $12H^4\epsilon^2$. The following analysis verifies explicitly that the inflationary branch lies on a regular Hamiltonian sheet.

Adopt the ADM variables $(N,N^i,\gamma_{ij})$ and define
\begin{equation}
 V_C\equiv N^{-1}(\dot C-N^iD_iC),
 \qquad
 V_\chi\equiv N^{-1}(\dot\chi-N^iD_i\chi),
 \qquad
 K_{ij}=\frac1{2N}(\dot\gamma_{ij}-D_iN_j-D_jN_i).
 \label{eq:dirac_velocities}
\end{equation}
After the Gibbons--Hawking--York term and the scalar--Gauss--Bonnet Dirichlet boundary term are included, the action contains $K_{ij}$, $V_C$, and $V_\chi$ but no $\dot K_{ij}$, $\dot N$, or $\dot N^i$ \cite{bib0005}.  The $C$--$\chi$ kinetic term gives
\begin{equation}
 P_C=\frac{1}{\sqrt\gamma}\frac{\partial L}{\partial V_C}=V_\chi,
 \qquad
 P_\chi=V_C+P_\chi^{\rm GB}(K,D\chi),
 \label{eq:dirac_scalar_momenta}
\end{equation}
and the Gauss--Bonnet term makes the metric momentum nonlinear in $K_{ij}$.

It is useful to test the Legendre map itself.  Let $U^I=(K^i{}_j,V_C,V_\chi)$ and let $\Pi_I$ denote the corresponding momentum densities divided by $\sqrt\gamma$.  In four spacetime dimensions the Jacobian has the exact block form
\begin{equation}
 \mathbb J^I{}_J\equiv\frac{\partial\Pi_I}{\partial U^J}
 =\begin{pmatrix}
 \mathfrak J&0&\mathfrak B\\
 0&0&1\\
 \mathfrak B^T&1&0
 \end{pmatrix},
 \qquad
 \det\mathbb J=-\det\mathfrak J.
 \label{eq:full_legendre_jacobian}
\end{equation}
The equality for the determinant follows by a Schur complement and is independent of the metric--$V_\chi$ mixing vector $\mathfrak B$.  Thus the response cross kinetic term does not create a new singular Legendre sheet.

Using the generalized Kronecker symbol
$\delta^{i_1\cdots i_n}_{j_1\cdots j_n}=n!\delta^{[i_1}_{j_1}\cdots\delta^{i_n]}_{j_n}$,
the metric block, normalized to the Einstein kinetic coefficient, is the $D=4$ specialization of the exact Einstein--scalar--Gauss--Bonnet ADM Jacobian \cite{bib0005}:
\begin{equation}
 \widehat{\mathfrak J}^{\,i}{}_{j}{}^{b}{}_{a}
 =\delta^{ib}_{ja}
 -\frac{8\ell_*}{M_{\rm Pl}^2}
 \delta^{ibk}_{jal}
 \left(D^lD_k\chi-V_\chi K^l{}_k\right).
 \label{eq:metric_legendre_block}
\end{equation}
An overall nonzero sign associated with choosing $\gamma_{ij}$ or $\gamma^{ij}$ as canonical coordinate does not affect its rank.  Eq.~\eqref{eq:metric_legendre_block} is exact for the linear coupling $f(\chi)=\ell_*\chi$; the rank-four Lovelock term vanishes identically on a three-dimensional hypersurface.

On the homogeneous inflationary branch,
\begin{equation}
 D_i\chi=0,
 \qquad D_iD_j\chi=0,
 \qquad K^i{}_j=H\delta^i{}_j,
 \label{eq:FLRW_legendre_specialization}
\end{equation}
so the six metric eigenvalues are proportional to a single factor
\begin{equation}
 Q_T\equiv1+\frac{8\ell_*H\dot\chi}{M_{\rm Pl}^2}.
 \label{eq:QT_regular_factor}
\end{equation}
There are five traceless eigenvalues proportional to $Q_T$ and one trace eigenvalue proportional to $-2Q_T$.  Therefore, up to the fixed nonzero normalization of the DeWitt supermetric,
\begin{equation}
 |\det\mathfrak J|=2M_{\rm Pl}^{12}|Q_T|^6,
 \qquad
 \det\mathbb J\ne0\quad\Longleftrightarrow\quad Q_T\ne0.
 \label{eq:legendre_determinant_FLRW}
\end{equation}
This identifies the only possible Legendre singularity of the homogeneous branch.

The numerical solution in Section~\ref{subsec:explicit_activation} gives, in the dimensionless units used there,
\begin{equation}
 \max_N|8\ell_*H\dot\chi|=935.4184\,H_*^2.
 \label{eq:QT_numeric_coefficient}
\end{equation}
At the conservative inflationary scale $H_*/M_{\rm Pl}\le10^{-5}$,
\begin{equation}
 |Q_T-1|<9.36\times10^{-8},
 \qquad
 Q_T>0.999999906,
 \qquad
 \det\mathbb J\ne0
 \label{eq:QT_numeric_bound}
\end{equation}
throughout the complete activation and deactivation history.  Thus the branch used in the observable calculation never approaches the multivalued-Hamiltonian surface discussed for strong-coupling Einstein--scalar--Gauss--Bonnet solutions \cite{bib0005}.

The Dirac--Bergmann chain starts from the fact that $N$ and $N^i$ carry no velocities, so
\begin{equation}
 \pi_N\approx0,
 \qquad
 \pi_i\approx0
 \label{eq:primary_gravity_constraints}
\end{equation}
are four primary constraints.  On the regular Legendre branch, the canonical Hamiltonian is linear in lapse and shift,
\begin{equation}
 H_C=\int d^3x\left(N\mathcal H_\perp+N^i\mathcal H_i\right)+H_{\rm alg},
 \label{eq:canonical_Hamiltonian_dirac}
\end{equation}
where $H_{\rm alg}$ contains the algebraic closure sector.  Preservation of the primary constraints yields
\begin{equation}
 \dot\pi_N=\{\pi_N,H_T\}=-\mathcal H_\perp\approx0,
 \qquad
 \dot\pi_i=\{\pi_i,H_T\}=-\mathcal H_i\approx0,
 \label{eq:secondary_gravity_constraints}
\end{equation}
with no tertiary constraint on the regular sheet.  Diffeomorphism invariance gives the standard smeared hypersurface-deformation algebra
\begin{align}
 \{H[\vec N],H[\vec M]\}&=H[[\vec N,\vec M]],
 \label{eq:dirac_alg1}\\
 \{H_\perp[N],H[\vec M]\}&=-H_\perp[\mathcal L_{\vec M}N],
 \label{eq:dirac_alg2}\\
 \{H_\perp[N],H_\perp[M]\}&=
 H\!\left[\gamma^{ij}(N\partial_jM-M\partial_jN)\right].
 \label{eq:dirac_alg3}
\end{align}
Hence $(\pi_N,\pi_i,\mathcal H_\perp,\mathcal H_i)$ are eight first-class constraints.

Before imposing the retarded restriction, the local $(g_{\mu\nu},C,\chi)$ phase space has dimension
\begin{equation}
 \dim\Gamma_{gC\chi}=2(1+3+6+1+1)=24.
 \label{eq:local_phase_dimension}
\end{equation}
The eight first-class constraints remove sixteen dimensions, leaving
\begin{equation}
 N_{gC\chi}^{\rm local}=\frac{24-2\times8}{2}=4
 =2_{\rm graviton}+2_{C,\chi}.
 \label{eq:local_dof_count}
\end{equation}
This reproduces the two localized response modes diagnosed in Section~\ref{subsec:retarded_phase_space}; they are not hidden by the gravitational constraint algebra.  Proposition~\ref{prop:retarded_reduction} subsequently removes their homogeneous Cauchy data from the physical retarded theory.

The five complementary components $(A,B_i,\Theta)$ have no velocities.  For example
\begin{equation}
 \pi_A\approx0,
 \qquad
 \Phi_A\equiv M_A^2A\approx0,
 \qquad
 \{\pi_A(x),\Phi_A(y)\}=-M_A^2\delta^{(3)}(x-y)\ne0,
 \label{eq:closure_second_class}
\end{equation}
with analogous second-class pairs for the three $B_i$ components and $\Theta$.  They remove all five non-STF parent-tensor components.  The inflaton contributes one physical scalar mode and the dressed STF field contributes five.  The final independent configuration-space content on the regular retarded inflationary branch is therefore
\begin{equation}
 {N_{\rm phys}=2_{\rm graviton}+1_{\rm inflaton}+5_Q=8.}
 \label{eq:total_dof_count}
\end{equation}
There is no canonical pair associated with $\dot K_{ij}$, the Legendre map is nonsingular on the entire benchmark trajectory, and the only negative-sign scalar combination of the localization is excluded by the retarded phase-space definition.  These three statements address, respectively, higher-derivative Ostrogradsky modes, Hamiltonian branch singularities, and the localized response ghost.

\section{ADM Constraint System}
\label{app:adm}

\subsection{Exact projections}\label{Xsec27-E.1}

Write
\begin{equation}
 ds^2=-N^2dt^2+\gamma_{ij}(dx^i+N^idt)(dx^j+N^jdt),
 \label{Xeqn280-E.1}
\end{equation}
with unit normal $n_\mu=(-N,0)$ and extrinsic curvature
\begin{equation}
 K_{ij}=\frac{1}{2N}
 \left(\dot\gamma_{ij}-D_iN_j-D_jN_i\right).
 \label{eq:K_ADM_section6}
\end{equation}
Define the combined memory--history projections
\begin{equation}
 \mathcal E_{\rm mem}=n^\mu n^\nu T_{\mu\nu}^{\rm mem},
 \qquad
 \mathcal P_{i,{\rm mem}}=-\gamma_i{}^\mu n^\nu T_{\mu\nu}^{\rm mem}.
 \label{eq:app_ADM_sources}
\end{equation}
The Gauss--Codazzi relations give the exact constraints
\begin{align}
& {}^{(3)}R+K^2-K_{ij}K^{ij}
 =\frac{2}{M_{\rm Pl}^2}
 \left[
 \frac12\Pi_\phi^2+\frac12D_i\phi D^i\phi+U(\phi)
 +\mathcal E_{\rm mem}
 \right],
 \label{eq:exact_H_constraint_total_section6}\\
& D_j(K^j{}_i-\delta^j{}_iK)
 =\frac{1}{M_{\rm Pl}^2}
 \left[\Pi_\phi D_i\phi+\mathcal P_{i,{\rm mem}}\right].
 \label{eq:exact_M_constraint_total_section6}
\end{align}
The geometric left-hand sides are exactly those of General Relativity.  All model dependence enters through projections of the conserved source, which is why lapse and shift remain nondynamical.

For the homogeneous background, $K^i{}_j=H\delta^i{}_j$ and the Hamiltonian constraint becomes
\begin{equation}
 3M_{\rm Pl}^2H^2
 =\frac12\dot{\bar\phi}^{\,2}+U(\bar\phi)
 +\bar{\mathcal E}_{\rm mem}.
 \label{eq:background_Hamiltonian_section6}
\end{equation}

\subsection{Linear scalar and transverse constraints}\label{Xsec28-E.2}

In comoving gauge,
\begin{equation}
 N=1+\alpha,
 \qquad
 N_i=\partial_i\beta+N_i^T,
 \qquad
 \gamma_{ij}=a^2e^{2\zeta}\delta_{ij},
 \qquad
 \partial_iN_i^T=0.
 \label{Xeqn286-E.7}
\end{equation}
To first order,
\begin{align}
 {}^{(3)}R&=-4a^{-2}\partial^2\zeta,\\
 K^i{}_j&=H\delta^i{}_j+(\dot\zeta-H\alpha)\delta^i{}_j
 -a^{-2}\partial^i\partial_j\beta
 -\frac12a^{-2}(\partial^iN_j^T+\partial_jN^{Ti}).
 \end{align}
Decompose the memory momentum as
\begin{equation}
 \delta\mathcal P_{i,{\rm mem}}
 =\partial_i\mathcal P_{\rm mem}
 +\mathcal P_{i,{\rm mem}}^T.
 \label{Xeqn287-E.10}
\end{equation}
The momentum and Hamiltonian constraints give the compact sourced solution
\begin{align}
 \alpha&=\frac{\dot\zeta}{H}
 +\frac{\mathcal P_{\rm mem}}{2M_{\rm Pl}^2H},
 \label{eq:app_alpha_NT}\\
 \frac{\partial^2\beta}{a^2}
 &=-\frac{1}{H}\frac{\partial^2\zeta}{a^2}
 +\frac{\dot{\bar\phi}^{\,2}}{2M_{\rm Pl}^2H^2}\dot\zeta
 -\frac{\delta\mathcal E_{\rm mem}}{2M_{\rm Pl}^2H}
 +\left[-\frac{3}{2M_{\rm Pl}^2}
 +\frac{\dot{\bar\phi}^{\,2}}{4M_{\rm Pl}^4H^2}\right]
 \mathcal P_{\rm mem},
 \label{eq:app_beta_sourced}\\
 N_i^T&=-\frac{2a^2}{M_{\rm Pl}^2}\partial^{-2}
 \mathcal P_{i,{\rm mem}}^T.
 \label{eq:app_NT_sourced}
\end{align}
These expressions follow directly from subtracting the background constraint before solving for the lapse and shift.  No time derivative of a source is needed; source evolution is constrained separately by the exchange identity.

In the complete coupled calculation the memory sources in \eqref{eq:app_alpha_NT} and \eqref{eq:app_beta_sourced} are \emph{not} set to zero.  For example, the cross-scalar part of the linear momentum potential is
\begin{equation}
 \mathcal P_{C\chi}^{(1)}=-(\dot{\bar\chi}\,c+\dot{\bar C}\,x),
 \label{eq:response_momentum_potential}
\end{equation}
and the Gauss--Bonnet part follows exactly from \eqref{eq:GB_compact_stress}.  The STF contribution is obtained from the metric variation of \eqref{eq:main_SDelta}.  These sourced constraints are substituted into the full Hessian before any propagating field is eliminated, as summarized by the exact Schur complement \eqref{eq:ADM_schur_exact}. Only after the retarded substitution \eqref{eq:c_ret_pert} and \eqref{eq:x_ret_pert} and expansion to first order in $b$ does the direct anisotropic $\zeta\zeta$ block reduce to the usual comoving-gauge lapse and shift; the remaining difference contributes at $O(b\Omega_{\rm mem})$ or $O(b^2)$.

\section{Quadratic Scalar Action and Finite-Window Kernel}
\label{app:quadratic_kernel}

\subsection{Constraint reduction and tracefree curvature coupling}\label{Xsec29-F.1}

The activated spatial tensor is
\begin{equation}
 \Delta_{ij}^{\rm aniso}=D(\bar C)
 \left(n_in_j-\frac13\gamma_{ij}\right),
 \qquad
 \Delta_{00}^{\rm aniso}=\Delta_{0i}^{\rm aniso}=0,
 \label{eq:Delta_aniso_section7}
\end{equation}
with the constant comoving projector
\begin{equation}
 P^{ij}=\hat n^i\hat n^j-\frac13\delta^{ij},
 \qquad \delta_{ij}P^{ij}=0.
 \label{eq:Pij_section7}
\end{equation}
The scalar constraint solution on this branch is
\begin{equation}
 \alpha=\frac{\dot\zeta}{H},
 \qquad
 \frac{\partial^2\beta}{a^2}
 =-\frac{1}{H}\frac{\partial^2\zeta}{a^2}
 +\frac{\dot{\bar\phi}^{\,2}}{2M_{\rm Pl}^2H^2}\dot\zeta.
 \label{eq:standard_scalar_constraints_section7}
\end{equation}
Insertion into the Einstein--inflaton action gives
\begin{equation}
 S_{\zeta,0}^{(2)}=M_{\rm Pl}^2
 \int dt\,d^3x\,a^3\epsilon
 \left[\dot\zeta^2-a^{-2}\partial_i\zeta\partial^i\zeta\right].
 \label{eq:S_zeta_standard_section7}
\end{equation}

For $\gamma_{ij}=a^2e^{2\zeta}\delta_{ij}$, the intrinsic spatial Ricci tensor is
\begin{equation}
 {}^{(3)}R_{ij}
 =-\partial_i\partial_j\zeta-\delta_{ij}\partial^2\zeta
 +\partial_i\zeta\partial_j\zeta
 -\delta_{ij}(\partial\zeta)^2.
 \label{eq:app_Rij_conformal}
\end{equation}
Contracting with $P^{ij}$ removes the trace terms.  The linear contribution is a boundary term, and the lapse-dependent term cancels after Eq.~\eqref{eq:standard_scalar_constraints_section7} is used.  The remaining quadratic term is
\begin{equation}
 S_{\Delta R}^{(2)}
 =2\eta\int dt\,d^3x\,aG(\bar C)D(\bar C)
 P^{ij}\partial_i\zeta\partial_j\zeta.
 \label{eq:app_aniso_action}
\end{equation}
Combining Eqs.~\eqref{eq:S_zeta_standard_section7} and \eqref{eq:app_aniso_action} yields
\begin{equation}
 S_\zeta^{(2)}=M_{\rm Pl}^2\int dt\,d^3x\,a^3\epsilon
 \left[\dot\zeta^2-a^{-2}(\delta^{ij}-bP^{ij})
 \partial_i\zeta\partial_j\zeta\right],
 \qquad
 b=\frac{2\eta G D}{M_{\rm Pl}^2\epsilon}.
 \label{eq:app_reduced_action}
\end{equation}
The eigenvalues of the gradient matrix are $1-2b/3$ parallel to $\hat n$ and $1+b/3$ in the transverse plane.  Positivity therefore gives
\begin{equation}
 -3<b<\frac32.
 \label{eq:app_gradient_interval}
\end{equation}
This is a gradient-stability condition.  The kinetic coefficient remains $M_{\rm Pl}^2\epsilon>0$, so no scalar ghost is generated at quadratic order.

\subsection{Endpoint response and infrared cancellation}\label{Xsec30-F.2}

With $v=z\zeta$, $z^2=2a^2M_{\rm Pl}^2\epsilon$, the perturbative interaction for a weak, approximately constant activation is proportional to
\begin{equation}
 H_I(\tau)=-\frac12b(\tau)
 \int\frac{d^3k}{(2\pi)^3}\,
 z^2k^2\left(\mu^2-\frac13\right)
 \zeta_{\mathbf k}\zeta_{-\mathbf k}.
 \label{eq:app_HI}
\end{equation}
Using the de Sitter mode function
\begin{equation}
 \zeta_k(\tau)=\frac{H_*}{2\sqrt{\epsilon_*}M_{\rm Pl}k^{3/2}}
 (1+ik\tau)e^{-ik\tau},
 \label{eq:app_mode}
\end{equation}
the in-in commutator reduces to an endpoint primitive.  For a top-hat activation, $\tau_i<\tau<\tau_f$, the quadrupole is
\begin{equation}
 g_*(k)=\mathcal A_{\rm mem}
 \left[\mathcal G(2k\tau_f)-\mathcal G(2k\tau_i)\right],
 \qquad
 \mathcal G(x)=\frac{4\sin x}{x}-\cos x.
 \label{eq:app_kernel}
\end{equation}
The constant term in the small-$x$ expansion,
\begin{equation}
 \mathcal G(x)=3-\frac{x^2}{6}+\frac{x^4}{120}+O(x^6),
 \label{Xeqn300-F.12}
\end{equation}
cancels between the two endpoints.  Therefore
\begin{equation}
 g_*(k)=-\frac{2}{3}\mathcal A_{\rm mem}k^2
 (\tau_f^2-\tau_i^2)+O(k^4).
 \label{eq:app_IR_kernel}
\end{equation}
The $k^2$ infrared behavior is a direct consequence of a finite episode: modes with wavelength much larger than both endpoint horizons cannot resolve the duration of the activation.  A smooth activation profile replaces the endpoint difference by a weighted time integral but preserves this cancellation when the profile has compact support or sufficiently rapid tails.

\section{BipoSH Projection, Likelihood, and Forecast Formulae}
\label{app:biposh_likelihood}

\subsection{Harmonic projection}\label{Xsec31-G.1}

Write the primordial spectrum as
\begin{equation}
 \mathcal P_\zeta(\mathbf k)=\mathcal P_0(k)
 \left[1+g_*(k)\left((\hat k\!\cdot\!\hat n)^2-\frac13\right)\right].
 \label{eq:app_primordial_quad}
\end{equation}
The addition theorem gives
\begin{equation}
 (\hat k\!\cdot\!\hat n)^2-\frac13
 =\frac{8\pi}{15}\sum_{M=-2}^{2}
 Y_{2M}(\hat k)Y_{2M}^*(\hat n).
 \label{eq:app_addition}
\end{equation}
For CMB fields $X\in\{T,E,B\}$,
\begin{equation}
 a_{\ell m}^X=4\pi(-i)^\ell
 \int\frac{d^3k}{(2\pi)^3}\,
 \zeta(\mathbf k)\Delta_\ell^X(k)Y_{\ell m}^*(\hat k).
 \label{Xeqn304-G.3}
\end{equation}
The angular integral of three spherical harmonics produces a Wigner-$3j$ symbol.  The anisotropic covariance can therefore be written
\begin{equation}
 \delta\langle a_{\ell m}^Xa_{\ell'm'}^{Y*}\rangle
 =\sum_M(-1)^{m'}A_{\ell\ell'}^{2M,XY}
 \begin{pmatrix}
 \ell&\ell'&2\\ m&-m'&M
 \end{pmatrix},
 \label{eq:app_BipoSH_definition}
\end{equation}
with
\begin{align}
 A_{\ell\ell'}^{2M,XY}={}&
 \frac{8\pi}{15}\,Y_{2M}^*(\hat n)
 (-i)^{\ell-\ell'}
 \sqrt{\frac{(2\ell+1)(2\ell'+1)5}{4\pi}}
 \begin{pmatrix}
 \ell&\ell'&2\\0&0&0
 \end{pmatrix}
 \mathcal I_{\ell\ell'}^{XY},\\
 \mathcal I_{\ell\ell'}^{XY}={}&4\pi
 \int d\ln k\,\mathcal P_0(k)g_*(k)
 \Delta_\ell^X(k)\Delta_{\ell'}^Y(k).
 \label{eq:app_radial_kernel}
\end{align}
The triangle and parity conditions imply $|\ell-\ell'|\le2$ and $\ell+\ell'+2$ even.  For scalar $T/E$ observables this leaves $\ell'=\ell$ and $\ell\pm2$.  The five $M$ components rotate as an irreducible $L=2$ object and reconstruct the unoriented axis $\hat n\sim-\hat n$.

\subsection{Likelihood and parameter mapping}\label{Xsec32-G.2}

Let the data covariance be
\begin{equation}
 \mathbf C(\vartheta)=\mathbf C^{(0)}+
 \sum_Mq_{2M}\mathbf C^{(2M)}(\tau_i,\tau_f),
 \qquad
 q_{2M}=\frac{8\pi}{15}\mathcal A_{\rm mem}Y_{2M}^*(\hat n).
 \label{eq:app_covariance_model}
\end{equation}
For Gaussian harmonic coefficients,
\begin{equation}
 -2\ln\mathcal L
 =\mathbf a^\dagger\mathbf C^{-1}\mathbf a+
 \ln\det\mathbf C+{\rm const}.
 \label{eq:app_likelihood}
\end{equation}
At weak anisotropy the Fisher matrix is
\begin{equation}
 F_{ab}=\frac12{\rm Tr}
 \left[\mathbf C_0^{-1}\mathbf C_{,a}
 \mathbf C_0^{-1}\mathbf C_{,b}\right].
 \label{eq:app_Fisher}
\end{equation}
The derivatives with respect to the endpoints follow directly from
\begin{equation}
 \frac{\partial g_*}{\partial\tau_f}
 =2k\mathcal A_{\rm mem}\mathcal G'(2k\tau_f),
 \qquad
 \frac{\partial g_*}{\partial\tau_i}
 =-2k\mathcal A_{\rm mem}\mathcal G'(2k\tau_i).
 \label{eq:app_endpoint_derivatives}
\end{equation}
A practical calculation precomputes the radial integrals for the amplitude and endpoint derivatives, constructs the allowed $\Delta\ell=0,2$ covariance blocks, and then applies the experiment's beam, mask, and anisotropic-noise response.

Finally,
\begin{equation}
 \eta G(C_0)D(C_0)=\mathcal A_{\rm mem}M_{\rm Pl}^2\epsilon_*.
 \label{eq:app_parameter_mapping}
\end{equation}
Thus a BipoSH limit constrains a definite covariant product rather than a free angular template.  The endpoint parameters carry information about the activation history, while the five $q_{2M}$ coefficients determine the amplitude and orientation of the tensor phase.

\end{document}